\documentclass[12pt,letterpaper]{article}
\pdfoutput=1

\usepackage[dvipsnames]{xcolor}
\usepackage{amsmath,amssymb,calc, amsthm,bbm, epsfig,psfrag, mathtools}
\usepackage{graphicx, enumerate}
\usepackage[numbers,sort&compress]{natbib}

\usepackage{tensor}
\usepackage{mathtools}
\usepackage{caption}
\usepackage{subcaption}
\usepackage{tikz-cd}
\usepackage{mathrsfs}
\usepackage[normalem]{ulem}
\usepackage{dsfont}

\DeclareMathAlphabet{\mathbcal}{U}{BOONDOX-cal}{m}{n}
\DeclareMathAlphabet{\mathbscriptcal}{U}{boondox-script}{m}{n}
\DeclareMathAlphabet{\mathdcal}{U}{dutchcal}{m}{n}
\DeclareMathAlphabet{\matheucal}{U}{eucal}{m}{n}

\usepackage[utf8]{inputenc} 

\newcommand{\Comment}[1]{{}}
\definecolor{darkblue}{rgb}{0.15,0.35,0.55}
\definecolor{reddish}{rgb}{0.65, 0.2, 0.2}

\usepackage[linktocpage=true]{hyperref}
\hypersetup{
colorlinks=true,
citecolor=darkblue,
linkcolor=reddish,
urlcolor=darkblue,
pdfauthor={},
pdftitle={},
pdfsubject={}
}

\PassOptionsToPackage{linecolor=blue,backgroundcolor=blue!25,bordercolor=blue,textsize=scriptsize}{todonotes}
\usepackage{todonotes}

\makeatletter
\renewcommand\section{\@startsection {section}{1}{\z@}%
                                   {-3.5ex \@plus -1ex \@minus -.2ex}
                                   {2.3ex \@plus.2ex}%
                                   {\normalfont\large\bfseries}}
\renewcommand\subsection{\@startsection{subsection}{2}{\z@}%
                                     {-3.25ex\@plus -1ex \@minus -.2ex}%
                                     {1.5ex \@plus .2ex}%
                                     {\normalfont\bfseries}}
\makeatother

\let\non\nonumber

\def\bea#1\eea{\begin{align}#1\end{align}}

\def\bes #1\ees{\begin{split}#1\end{split}}

\newcommand{\be}{\begin{equation}}
\newcommand{\ee}{\end{equation}}

\newcommand{\bma}{\begin{pmatrix}}
\newcommand{\ema}{\end{pmatrix}}

\newcommand{\hlf}{\frac{1}{2}}

\renewcommand{\ss}{\,\,\,}

\newcommand{\fs}{\mathcal{F}}

\let\x=\xi

\def\x{\xi}

\newcommand{\com}[2]{{ \left[ #1, #2 \right] }}

\newcommand{\p}{\partial}

\def\com#1#2{{ \left[ #1, #2 \right] }}

\newcommand{\C}[1]{(\ref{#1})}

\def\IZ{\relax\ifmmode\mathchoice
{\hbox{\cmss Z\kern-.4em Z}}{\hbox{\cmss Z\kern-.4em Z}}
{\lower.9pt\hbox{\cmsss Z\kern-.4em Z}} {\lower1.2pt\hbox{\cmsss
Z\kern-.4em Z}}\else{\cmss Z\kern-.4em Z}\fi}
\def\IR{\relax{\rm I\kern-.18em R}}

\def\one{{\hbox{ 1\kern-.8mm l}}}

\newlength{\bredde}
\def\slash#1{\settowidth{\bredde}{$#1$}\ifmmode\,\raisebox{.15ex}{/}
\hspace*{-\bredde} #1\else$\,\raisebox{.15ex}{/}\hspace*{-\bredde}
#1$\fi}

\newsavebox{\zzzbar}
\sbox{\zzzbar}
  {\setlength{\unitlength}{0.9em}
  \begin{picture}(0.6,0.7)
  \thinlines
  \put(0,0){\line(1,0){0.6}}
  \put(0,0.75){\line(1,0){0.575}}
  \multiput(0,0)(0.0125,0.025){30}{\rule{0.3pt}{0.3pt}}
  \multiput(0.2,0)(0.0125,0.025){30}{\rule{0.3pt}{0.3pt}}
  \put(0,0.75){\line(0,-1){0.15}}
  \put(0.015,0.75){\line(0,-1){0.1}}
  \put(0.03,0.75){\line(0,-1){0.075}}
  \put(0.045,0.75){\line(0,-1){0.05}}
  \put(0.05,0.75){\line(0,-1){0.025}}
  \put(0.6,0){\line(0,1){0.15}}
  \put(0.585,0){\line(0,1){0.1}}
  \put(0.57,0){\line(0,1){0.075}}
  \put(0.555,0){\line(0,1){0.05}}
  \put(0.55,0){\line(0,1){0.025}}
  \end{picture}}

\newfont{\goth}{ygoth.tfm scaled 1200}                   

 \numberwithin{equation}{section}

\def\1{{(1)}}
\def\2{{(2)}}
\def\3{{(3)}}

\newcommand{\ul}{\underline}

\newcommand{\tdelta}{\tilde{\delta}}
\newcommand{\tepsilon}{\tilde{\epsilon}}
\newcommand{\tpsi}{\tilde{\psi}}

\begin{document}
\begin{titlepage}

\begin{center}

July 2, 2026
\hfill         \phantom{xxx}  EFI-24-3

\vskip 2 cm {\Large \bf Supersymmetry and Nonreciprocity}
\vskip 1.25 cm {\bf Savdeep Sethi\footnote{\href{mailto:EMAIL1}{\texttt{sethi@uchicago.edu}}} and Gabriel Artur Weiderpass\footnote{\href{mailto:EMAIL1}{\texttt{gaweiderpass@uchicago.edu}}}}\non\\
\vskip 0.2 cm

{\it Leinweber Institute for Theoretical Physics \& Enrico Fermi Institute \\ \& Kadanoff Center for Theoretical Physics, \\ University of Chicago, Chicago, IL 60637, USA}
\vskip 0.2 cm

\vskip 0.2 cm

\end{center}
\vskip 1.5 cm

\begin{abstract}
\baselineskip=18pt\textbf{}

Nonreciprocal theories are used to model a broad array of non-equilibrium phenomena found in nature ranging from biological systems like networks of neurons to the behavior of overflowing water fountains. This includes systems broadly classified as active matter systems. We show that the stochastic theories which describe nonreciprocal interactions can be mapped into quantum field theories described by a supersymmetric action with a single supercharge. The theories are generically non-Hermitian. This generalizes the past work of Parisi and Sourlas on reciprocal theories, which model systems with interactions derived from potentials.

\end{abstract}

\end{titlepage}

\tableofcontents

\section{Introduction} \label{intro}

\subsubsection*{\ul{\it What is nonreciprocity?}}

While the microscopic fundamental forces seen in nature are conservative, there is a vast range of physical phenomena modeled by forces which are not conservative~\cite{aguilera_unifying_2021,ivlev_statistical_2015,you_nonreciprocity_2020,brauns_nonreciprocal_2024,frohoff-hulsmann_nonreciprocal_2023,saha_scalar_2020,liu_non-reciprocal_2023,hanai_critical_2020,avni2025nonreciprocalPRL,Avni2025DynPhase,liu2025universal,zelle2024universal,young2020nonequilibrium,loos_irreversibility_2020,du_hidden_2024,marchetti_hydrodynamics_2013,hanai_non-hermitian_2019,rajeev_ising_2024,seara_non-reciprocal_2023,hanai_nonreciprocal_2024,han_coupled_2024,godreche_nonequilibrium_2009,godreche_dynamics_2011,godreche_rates_2013,godreche_dynamics_2014,godreche_dynamics_2015,godreche_generic_2017,godreche_freezing_2018,Weiderpass2025PRE,Weis-Fruchart2023Exceptional,Weis-Hanai2025Generalized,ashida_non-hermitian_2020}. Newton's third law of motion can be violated
in such non-equilibrium systems,
which include active matter  and soft matter systems \cite{brandenbourger_non-reciprocal_2019,veenstra_non-reciprocal_2024,brandenbourger_non-reciprocal_2024,ghatak_observation_2020,helbig_generalized_2020,kotwal_active_2021,hofmann_chiral_2019,gupta_active_2022,tan_odd_2022,shankar_topological_2022,fruchart_odd_2023,colen_interpreting_2024,scheibner_odd_2020,poncet_when_2022,rosa_dynamics_2020,scheibner_non-hermitian_2020,coulais_topology_2021,dinelli_non-reciprocity_2023,Guillet2025PNAS,van2024soft},  and open quantum systems \cite{metelmann_nonreciprocal_2015,mcdonald_exponentially-enhanced_2020,clerk_introduction_2022,chiacchio_nonreciprocal_2023,bergholtz_exceptional_2021,begg_quantum_2024,belyansky2025nonreciprocal,jachinowski2025spinonly}. 
The interactions are typically nonreciprocal: widget $A$ exerts a force on widget $B$, but the reaction is not equal and opposite. For some recent reviews, see~\cite{fruchart2026nonreciprocal,fruchart_non-reciprocal_2021, Clerk-IntroNR-2022, sounas_non-reciprocal_2017, Asadchy-NRandEM2020} which cover aspects of nonreciprocity in soft matter systems, in open quantum systems, in acoustics and in electromagnetism, respectively.
Rather than discussing generalities, we will illustrate nonreciprocity with a simple example. Imagine two linear springs satisfying Hooke's law with forces, 
\begin{align}
 \begin{pmatrix}
        f_x \\
        f_y
    \end{pmatrix}  = \begin{pmatrix}
        -k_x & 0 \\ 0 & -k_y
    \end{pmatrix} \begin{pmatrix}
        x \\ y
    \end{pmatrix}\, ,
\end{align}
and introduce couplings so that
\begin{align} \label{coupledsprings}
 \begin{pmatrix}
        f_x \\
        f_y
    \end{pmatrix}  = \begin{pmatrix}
        -k_x & a \\ b & -k_y
    \end{pmatrix} \begin{pmatrix}
        x \\ y
    \end{pmatrix} \, .
\end{align}
For $a \neq b$, this force is not the gradient of a potential energy and the theory is now nonreciprocal. 

To model systems of this type, which are typically far from equilibrium with sizable fluctuations, we can use a stochastic dynamical system. Without including any stochastic noise, the system is deterministic. Even in this purely classical case, there are rich dynamical phenomena that are poorly understood, seen, for instance, in models of machine learning and neuroscience; for a recent analysis, see~\cite{PhysRevX.12.011011}. The goal of this work is to map nonreciprocal stochastic dynamical systems into quantum field theories, which will typically be non-Hermitian. For reciprocal systems that have a potential energy, this mapping was described $45$ years ago by Parisi and Sourlas~\cite{Parisi-Soulars-1979, Parisi-Soulars-1982}.  In that case, the resulting field theory is Hermitian and, somewhat surprisingly, also supersymmetric with one complex supercharge. For recent reviews, applications, and discussions of the breakdown of Parisi-Sourlas supersymmetry and dimensional reduction, see \cite{rychkov2023lectures,kaviraj2020random,kaviraj2021random,kaviraj2022parisi,LeFloch:2025qjc,Tarjus-2025,Tarjus-2020,Tarjus-2018}.

Repeating the procedure of Parisi and Sourlas in a straightforward way does not give a supersymmetric action in the nonreciprocal case. The supersymmetry is broken by non-conservative forces, although a nilpotent BRST charge  still survives. 
For a review of stochastic systems and supersymmetry, which summarizes the current state of the art, see~\cite{Ovchinnikov:2015rkh}. An argument that suggests supersymmetry might exist even in the nonreciprocal case can also be found in~\cite{Ovchinnikov:2015rkh}; for applications of this work, see~\cite{ovchinnikov2016kinematic,ovchinnikov2018supersymmetric,westerkamp2021dynamical,ovchinnikov2021toward,zhai2021witten,Ovchinnikov-TFT-dynamical-systems,ovchinnikov2025topological}. The supercharge is constructed implicitly in terms of the exact energy spectrum of the theory.\footnote{We would like to thank Igor Ovchinnikov for a helpful discussion about this issue.} What is missing is an explicit supersymmetric action in the spirit of Parisi and Sourlas. We will show that such a supersymmetric action with, generically, one non-Hermitian supersymmetry exists even with nonreciprocal interactions. For an interesting different approach to supersymmetry and nonreciprocity, see~\cite{marguet_supersymmetries_2021}.

\subsubsection*{\ul{\it Outline \& Conclusions}}

In Section \ref{MSR} we begin by reviewing the MSR map which formulates a stochastic differential equation in terms of a quantum theory. For the case of a stochastic ODE, the equivalent quantum theory is quantum mechanics. If the theory is reciprocal then the quantum theory has one complex supersymmetry. If the theory is nonreciprocal, the supersymmetry is explicitly broken and there is only a nilpotent BRST charge \cite{Zinn-Justin2021qftc}. There is a rather beautiful analogy between the way nonreciprocity appears in the quantum theory and the coupling of a particle to a background magnetic field.

We then proceed in section \ref{Sec-SUSY-NR} to formulate a quantum description of the nonreciprocal case which possesses a manifest $N=1$ supersymmetry with one supercharge. The action is given in superspace in \C{(0,1)Inspired-QM} and in components in \C{S-New-No-Auxiliary}. The theory is generically non-Hermitian, which allows for a wider range of physical phenomena like the appearance of exceptional points.

In Section \ref{Sec-Ward} we derive the Ward identities associated with these symmetries. In the reciprocal Parisi-Sourlas formulation, the BRST charge enforces causality and probability conservation, while the second supercharge gives the fluctuation-dissipation theorem. In the nonreciprocal formulation, the single unbroken $N=1$ supersymmetry gives exact source-free identities even away from equilibrium; the usual fluctuation-dissipation violation is recast as a response-ghost condensate controlled by the nonreciprocal curvature.

In Section \ref{Sec-SPDE} we generalize the discussion of reciprocal systems from ODEs to stochastic PDEs. The corresponding quantum field theories again have one complex supersymmetry. We discuss two examples: the first is the stochastic heat equation which maps to a supersymmetric Lifshitz theory. The second example is the stochastic equation describing the evolution towards equilibrium of the Ising model, which maps to an interacting supersymmetric Lifshitz theory. 

In Section \ref{Sec-NR-SPDE} we apply the technology developed in Section \ref{Sec-SUSY-NR} to nonreciprocal stochastic PDEs describing models with non-conserved order parameters. These models have broad applicability and lately have been used to model nonreciprocal interactions in soft, active and condensed matter physics~\cite{fruchart_non-reciprocal_2021,liu_non-reciprocal_2023,hanai_critical_2020,avni2025nonreciprocalPRL,Avni2025DynPhase,liu2025universal,gupta_active_2022,zelle2024universal,young2020nonequilibrium}. We show that the quantum field theories corresponding to nonreciprocal stochastic PDEs can be formulated with a manifest $N=1$ non-Hermitian supersymmetry. The action is again given in superspace in \C{FT-Action-Superspace} and in components in \C{FT-Action}. We again discuss three particular cases of stochastic PDEs: the first is a nonreciprocal extension of the stochastic $O(2)$ model. The second is two stochastic heat equations coupled nonreciprocally. The final model is the continuum description of two kinetic Ising models with a nonreciprocal diffusive term.  Section \ref{Sec-Discussion} is a summary of our results.

Lastly in Appendix \ref{App-Ito-Stratonovich}, we describe the standard approach to regularizing stochastic ODEs, which depends on a parameter $\alpha$. Appendix \ref{App-Grassmanian-Integrals} has useful identities about Grassmann integration used in the bulk of the text, while Appendix \ref{symplectic} has a discussion of symplectic quantization.

There are many basic questions to address and directions to explore in the future. The supersymmetric quantum theory superficially contains more information than the stochastic model. Which states in the quantum Hilbert space admit a stochastic interpretation? With reciprocity, the quantum theory has a conserved fermion number. With nonreciprocity, the supersymmetric formulation of the theory does not appear to possess a conserved fermion number. What does this imply about the structure of eigenstates? Can supersymmetry, via quantities like the index and methods like localization, help us learn about the long time behavior of nonreciprocal theories? About possible critical behavior? About the physics of exceptional points? It would also be interesting to extend this analysis to disordered systems discussed from a supersymmetric perspective recently in~\cite{rychkov2023lectures}. Lastly, there are also topological mechanical systems that admit supersymmetric descriptions, where a similar reduction in supersymmetry from a kind of nonreciprocity has been seen~\cite{Attig_2019,upadhyaya2020nuts}. 

\section{Stochastic ODEs and Supersymmetric Quantum Mechanics} \label{MSR}

Consider a set of variables $\phi^i(t)$ with $i=1,...,n$ that describe a dynamical system with $n$-dimensional target space $M$.  Take a vector field $f^i(\phi)$ on $M$ which we will assume has no additional time-dependence beyond the dependence on the fields $\phi(t)$. This is essentially the force\footnote{Calling $f^i$ the force is more intuitive if the dynamical system is viewed as a system in an overdamped limit where accelerations are dropped. For example, a system like $\frac{1}{2\gamma}\ddot{\phi}^i + \dot{\phi}^i = f^i$ in the limit where $\gamma$ is very large and suppresses the acceleration term. With this interpretation in mind, we will refer to $f^i$ as the force.} but, crucially, we will not assume it is the gradient of any scalar function. The vector field $f^i$ describes a flow on $M$ and can be used to define a deterministic dynamical system:
\begin{align} \label{DynSys}
    \dot \phi^i =  f^i\,.
\end{align}
We can use this dynamical system to define a stochastic ordinary differential equation on this manifold. The stochastic ODE associated to the dynamical system (\ref{DynSys}) is called the Langevin equation and is given by\footnote{Sometimes the term Langevin equation is reserved for dynamics defined by a potential $f^i = \partial^i V$.}
\begin{align} \label{Langevin-General}
    \dot \phi^i =  f^i + \xi^i\,,
\end{align}
where $\xi^i$ is Gaussian white noise defined by a partition function,
\begin{align} \label{NoiseMeasure}
    Z = \int D\xi_i \exp \left(- \int  dt \ \frac{\xi_i\xi^i}{2\sigma} \right)\,,
\end{align}
and we normalize the Gaussian measure to give correlation functions:
\begin{align} \label{noisecorrel}
    \langle \xi^i(t) \rangle = 0\,, \qquad \langle \xi^i(t) \xi^j(t') \rangle = \sigma \delta^{ij} \delta(t-t')\,.
\end{align}
There are many generalizations of this structure, including multiplicative Gaussian white noise discussed in Appendix \ref{App-Ito-Stratonovich}, and non-Gaussian noise terms, which appear in numerous applications of stochastic differential equations \cite{gardiner2004handbook,kunita1990stochastic,hsu2002stochastic,Bruned2022geometric,Friz2020}. Here we focus on the simplest case of Gaussian white noise.
The $\sigma$ appearing in \C{noisecorrel} is the amplitude of the noise.
Even though the Langevin equation (\ref{Langevin-General}) can easily be generalized to curved Riemannian manifolds, see Section 34.9 of \cite{Zinn-Justin2021qftc} and \cite{Ovchinnikov:2015rkh,kunita1990stochastic,hsu2002stochastic,Bruned2022geometric}, it will be simpler and sufficient for our discussion to consider a flat target manifold. We therefore assume a flat Euclidean metric $g_{ij} = \delta_{ij}$ for $M$. The target manifold can still have a non-trivial topology if it contains a torus with some compact periodic directions. 

Defining the stochastic theory requires a choice of regularization of the noise term in \C{Langevin-General}. There are two popular approaches called Ito or Stratonovich regularization; see  Appendix \ref{App-Ito-Stratonovich} and~\cite{ezawa_fermions_1985,Damgaard:1987rr,gardiner2004handbook} for a discussion oriented toward physicists. In physical applications, the Stratonovich approach is typically used so the conventional rules of calculus can be employed. We will take the philosophy that the map into a path-integral description, given below, defines the regularized stochastic system. 

\subsection{Formulating stochastic ODEs using the MSR path-integral}

To compute correlation functions in the stochastic theory, we solve equation (\ref{Langevin-General}) subject to initial conditions, $\phi^i(t_0) = \phi^{i}_0$,  while treating the noise $\xi^i(t)$ as an external source.  
We then average the correlation function over the noise,
\begin{align} \label{CorrSODE}
    \langle \phi^{i_1}_\xi(t_1) \ldots \phi^{i_n}_\xi(t_n) \rangle = \int D\xi^i(t') \, \phi^{i_1}_\xi(t_1) \ldots \phi^{i_n}_\xi(t_n) \exp \left(- \int  dt' \ \frac{\xi_i(t')^2}{2\sigma} \right)\,,
\end{align}
where $\phi^i_\xi(t)$ is formally a solution of \C{Langevin-General}. To proceed we follow the Martin-Siggia-Rose (MSR) procedure, which is basically the standard Faddeev-Popov trick used in gauge theory but applied to this stochastic system~\cite{Kamenev_2011,Altland_Simons_2010}.  

Denoting the classical equation of motion $E^i[\phi(t)] = \dot \phi^i(t) - f^i\big(\phi(t)\big)$, we insert the  Faddeev-Popov identity
\begin{align} \label{FP-identity}
     \mathds{1} = \int D\phi^i\, \delta\big(E^i - \xi^i\big) \text{Jac} \left(\frac{\delta E^i}{\delta \phi^j}\right)\,,
\end{align}
in the path integral (\ref{NoiseMeasure}). At this point, it is common to introduce a response field ${\hat\phi}^i$ and write the $\delta$-function as follows:
\begin{align} \label{MSR-path-integral-response-field}
    Z & = \int D\xi^i D \phi^i D{\hat\phi}^i \left|{\det}^{\prime} \left(\frac{\delta E^i} {\delta \phi^j}\right) \right| \exp \left(- \int  dt \left[ -i\hat\phi_i(E^i-\xi^i) + \frac{\xi_i \xi^i}{2\sigma}\right]\right)\,, \\
    & = \int D \phi^i D{\hat\phi}^i \left|{\det}^{\prime} \left(\frac{\delta E^i} {\delta \phi^j}\right) \right| \exp \left(- \int  dt \left[ -i\hat\phi_i \dot\phi^i + i \hat \phi_i f^i +  \frac{\sigma \hat\phi_i \hat\phi^i}{2}\right]\right)\,,
\end{align}
where we have integrated over the noise $\xi^i$. As a last step we integrate out the response field $\hat\phi^i$ and get
\begin{align} 
    Z
    & = \int D \phi^i \left|{\det}^{\prime} \left(\frac{\delta E^i} {\delta \phi^j}\right) \right| \exp \left(- \int  dt \left[ \frac{(\dot \phi^i - f^i)^2}{2\sigma}\right]\right)\,. \label{MSR-path-integral}
\end{align}
We used the fact that the Jacobian of the transformation $\xi(t) \mapsto \phi(t)$ is given by
\begin{align} \label{Functional-Jacobian}
    \text{Jac} \left(\frac{\delta E^i}{\delta \phi^j}\right) = \left|{\det}^{\prime} \left(\frac{\delta E^i}{\delta \phi^j}\right) \right|\,.
\end{align}
The strength of the noise, $\sigma$, sets $\hbar$ in the quantum theory. For simplicity, we will set $\sigma=1$ in our subsequent discussion. In the Jacobian (\ref{Functional-Jacobian}) the prime in the functional determinant indicates that we are omitting zero modes of $\frac{\delta E^i}{\delta \phi^j}$. If any zero modes should exist, they must be treated separately.\footnote{We do not typically expect any zero modes when specifying initial value data at a finite time. Examples with zero modes usually involve specifying boundary conditions at asymptotic boundaries like $t\rightarrow \pm \infty$. Such zero modes are often associated to symmetries like time translation invariance.}

\subsubsection*{\ul{\it Fokker-Planck Hamiltonian}}

As discussed in Appendix \ref{App-FP-derivation}, the conditional probability density $P(\phi,t|\phi_0,0)$ for the stochastic system satisfies the Fokker-Planck equation,
\begin{align} \label{FP-simple}
    \partial_t P = \partial_i  \left[ \left( \frac{\partial^i}{2} - f^i  \right)P\right]\,,
\end{align}
where $\partial_i = \frac{\partial}{\partial \phi^i}$. 
While not generally true for multiplicative noise, as shown in Appendix \ref{App-FP-derivation} the Fokker-Planck equation is independent of the regularization scheme, which depends on a parameter $\alpha$, for non-multiplicative Gaussian white noise. 

As discussed in Appendix \ref{Ito-Str-Path-int} when we define the path integral (\ref{MSR-path-integral-response-field}) on an interval $t\in[0,T]$ the functional determinant is a strictly positive function which depends on the regularization scheme and is given by
\begin{align}
    {\det}^{\prime} \left(\frac{\delta E^i} {\delta \phi^j}\right)  = \exp \left( - \alpha \int_0^T dt\, \partial_i f^i \right)\,,
\end{align}
with $\alpha = 0 $ and $\alpha=1/2$ corresponding to the Ito and Stratonovich regularizations, respectively. If the path integral \C{MSR-path-integral} is defined on an interval, we can therefore remove the absolute value in the Jacobian (\ref{Functional-Jacobian}).

Now choose the Ito regularization and notice that in (\ref{MSR-path-integral-response-field}) the response field $\hat \phi$ is the canonical momentum for that Euclidean path integral, which means that $\hat \phi_i = -i \partial_i$ in the operator formalism. Reading the Euclidean Hamiltonian from (\ref{MSR-path-integral-response-field}) gives,
\begin{align} \label{FP-Hamiltonian}
    H_{FP} = \frac{\hat \phi_i\hat\phi^i}{2} + i \hat \phi_i f^i\,.
\end{align}
We see that the Fokker-Planck equation (\ref{FP-Hamiltonian}) is just the Euclidean time Schr\"odinger equation $\partial_t P = - H_{FP} P$ for (\ref{FP-Hamiltonian}), which we call the Fokker-Planck Hamiltonian. For chosen boundary conditions on the interval $t\in[0,T]$, the path integrals (\ref{MSR-path-integral-response-field}) and (\ref{MSR-path-integral}) can be used to compute the conditional probability distribution via,
\begin{align}
    P(\phi,T|\phi_0,0) & = \langle \phi| e^{-H_{FP} T}|\phi_0\rangle\,, \\
    \label{MSR-path-integral-response-field-boundaries}
    & = \int\limits_{\phi(0)=\phi_0}^{\phi(T) = \phi } D \phi^i D{\hat\phi}^i \exp \left(- \int_0^T  dt \left[ -i\hat\phi_i \dot\phi^i + \frac{\hat\phi_i \hat\phi^i}{2} + i \hat \phi_i f^i \right]\right)\,, \\
    & = \int\limits_{\phi(0)=\phi_0}^{\phi(T) = \phi } D \phi^i \exp \left(- \int_0^T  dt \left[ \frac{\dot\phi^i\dot\phi_i}{2} - \dot \phi_i f^i + \frac{f^i f_i}{2} + \frac{\partial_i f^i}{2}\right]\right)\,. \label{MSR-path-integral-boundaries}
\end{align}
The last term in (\ref{MSR-path-integral-boundaries}) appears after integrating out $\hat \phi_i $. When expanding the square $(\phi^i-f^i)^2$, we must remember that  using equation (\ref{alpha-regularized-integral}) for Ito regularization gives
\begin{align}
    \int_0^T dt\, \dot\phi_i \circ_{0} f^i = \int_0^T dt \left[ \dot\phi_i f^i - \frac{\partial_if^i}{2} \right]\,,
\end{align}
where $\circ_{0}$ means that the integral on the left-hand side is the Ito integral. Even though we used Ito regularization in our argument here,  we would have arrived at the same Fokker-Planck Hamiltonian \C{FP-Hamiltonian} and path integral (\ref{MSR-path-integral-boundaries}) regardless of $\alpha$, as described in Appendix \ref{App-FP-derivation} and \ref{Ito-Str-Path-int}.

\subsubsection*{\ul{\it Introducing fermions}}

Our next step is to represent the functional determinant using fermions, 
\begin{align} \label{FunDetStandard}
    {\det}^{\prime}\left( E^{i}{}_{j}\right) = \int D\psi D\tilde \psi \exp \left( -\int dt \, \psi_i E^{i}{}_{j}  \tilde\psi^j \right)\,, 
\end{align}
where $E^{i}{}_{j} = \frac{\delta E^i}{\delta \phi^j} $ and the fermions $\psi$ and $\tilde\psi$ are independent real fermions. The determinant \C{FunDetStandard} is often expressed in terms of a pair of complex conjugate fermions rather than real fermions. For nonreciprocity, we found it more natural to use real fermions which provide an equivalent representation of the determinant; see Section~\ref{App-Real-vs-Complex} of Appendix \ref{App-Grassmanian-Integrals} for a discussion. This is the MSR representation of the partition function, 
\begin{align} \label{MSRpartition}
    Z = \int D\phi D\psi D\tilde \psi \exp \left(- \int dt \left[ \frac{(\dot \phi^i - f^i)^2}{2} + \tilde\psi^i \dot \psi_i + \tilde \psi^j\partial_j f^i \psi_i  \right]\right) \,,
\end{align}
with MSR action:
\begin{align} \label{MSR-Action}
    S_{MSR} = \int  dt \left[ \frac{(\dot \phi^i - f^i)^2}{2} + \tilde\psi^i  \dot \psi_i + \tilde \psi^j \partial_j f^i \psi_i \right]\,.
\end{align}
This action is quite special because the action is no more than quadratic in fermions for models with flat target spaces. 
There is one other point to note about \C{MSRpartition}. After integrating out the fermions, the path-integral can be used to evaluate $P(\phi,T|\phi_0,0) $ using \C{MSR-path-integral-response-field-boundaries}. 
However once we have the quantum formulation with fermions, we are free to study the corresponding quantum Hamiltonian along with its associated state space.
This is a larger structure than is visible in the purely bosonic stochastic theory though we might hope that interesting features of the stochastic model are encoded in the quantum structure.  

\subsection{Symmetries and the magic of reciprocity} \label{Sec-magic-of-reciprocity}

Let us review the symmetries of the MSR action \C{MSRpartition} with the specific goal of exhibiting supersymmetry when the model is reciprocal. As a preliminary, we need to discuss how to distinguish reciprocal models from nonreciprocal models. In this section, we will restrict our discussion of nonreciprocity to the zero momentum sector of the theory. More general non-equilibrium theories with interactions that are not gradients of a scalar functional, like KPZ, can also be treated in this framework.\footnote{We would like to thank Cheyne Weis for pointing this out to us.}

The dynamics is driven by the choice of vector field $f^i$. Since we have a metric $g_{ij} = \delta_{ij}$ on our target space, we can also study the form $f = f_i d\phi^i$. If the target space were compact we could then use Hodge decomposition to decompose this form
\begin{align}
   f = - dV - \delta U - \gamma\,,
\end{align}
where $d$ is the exterior derivative and  $\delta = (-1)^{n(p+1)+n}\star d \star $. The exact term determined by $V$ is reciprocal while the co-exact term $\delta U$ is nonreciprocal. The last term, $\gamma$, is harmonic. Whether it corresponds to a reciprocal or nonreciprocal interaction is more subtle to determine because it depends on how one wishes to define nonreciprocity physically. We will give some examples momentarily to illustrate this subtlety.

When the target space is non-compact, which is the typical situation, we need to be careful about the physical constraints we want to impose in determining what constitutes good potentials $(V,U)$. The unambiguous statement is that if $f$ is not closed, the system has nonreciprocal interactions. In terms of the vector field, 
\begin{align} \label{defA}
    f^i = -\partial^i V - A^i - \gamma^i\,,
\end{align}
a non-zero $A$ guarantees nonreciprocity where $A_i d\phi^i = \delta U$. 

As a first example, we can turn to the system \C{coupledsprings} taking $a=-b$ so
\begin{align}
    f = \frac{1}{2} d \left( k_x x^2 + k_y y^2\right) + a\left( y dx - x dy\right) \,,
\end{align}
so $A$ is non-zero and the system is nonreciprocal because $df = dA \neq 0$. On the other hand, take the target space $\mathbb{R}^2/\{0\}$ and consider the force
\begin{align} \label{harmonic-vec-fields}
    f= \frac{1 }{x^2 + y^2} \left(-y dx + x dy\right)\,.
\end{align}
In this case, $df=0$ but we still consider the system to be nonreciprocal because the forces are not equal and opposite. If we take polar coordinates $(r,\theta)$ for $\mathbb{R}^2/\{0\}$ then the force is formally exact $f=d\theta$ but $\theta$ is not a globally defined potential. This is a case where $f$ is harmonic. 
The upshot is that a sufficient, but not necessary condition, for nonreciprocity is $\fs=df$ with components $\fs_{ij} =\partial_i f_j - \partial_j f_i$ non-vanishing. We will primarily study this case when we consider nonreciprocal models.

In the reciprocal case with $A_i=0$, the existence of two supercharges was first pointed out by Parisi and Sourlas~\cite{Parisi-Soulars-1979,Parisi-Soulars-1982}; for a pedagogical discussion, see~\cite{Zinn-Justin2021qftc}. The first fermionic symmetry of \C{MSRpartition} acts as follows, 
\begin{align} \label{Q-In-MSR}
     \delta \phi^i = -\epsilon \tilde \psi^i\,, \qquad \delta \psi^i = \epsilon \left(\dot \phi^i - f^i\right)\,, \qquad \delta \tilde \psi^i = 0\,, 
\end{align}
where $\epsilon$ is a Grassmann parameter. The Noether procedure gives the supercharge,
\begin{align} \label{Q-class}
    Q = \tpsi^i \left( \dot\phi_i - f_i \right)\,. 
\end{align}
We will return to the operator approach later but for now we will continue to analyze the theory from a path-integral perspective. The symmetry \C{Q-In-MSR} is always present and it satisfies $\delta^2=0$ on-shell,  
\begin{align} 
     \delta_2 \delta_1 \phi^i = 0\,, \qquad \delta_2 \delta_1 \psi^i = 0\,, \qquad \delta_2 \delta_1 \tilde \psi^i = 0\,. \label{Qtilde-square-coord}
\end{align}
This is therefore a BRST symmetry rather than a supersymmetry. 

There is a second fermionic transformation of potential interest, 
\begin{align} \label{Qtilde-In-MSR}
    & \tdelta \phi^i = \tepsilon \psi^i\,, \qquad \tdelta \psi^i = 0\,, \qquad \tdelta \tilde \psi^i = \tepsilon \left(-\dot \phi^i - f^i\right)\,.
\end{align}
Acting on the action \C{MSR-Action} and keeping the Grassmann variable $\tepsilon(t)$ time-dependent enables us to read the charge which could potentially implement $\tdelta$. We get 
\begin{align} \label{Breaking-PS}
    \tdelta S_{MSR} = \int dt & \left[- \frac{d \tepsilon}{dt}\left[\psi_i\left(-\dot\phi^i - f^i\right)\right]-\frac{d}{dt}\Big(\tilde \epsilon\, 2 f^i \psi_i \Big)  \right. \cr & \ \, +\left. \tilde \epsilon \left( \left(\dot \phi^i - f^i\right) \psi^j -\frac{\psi^i \psi^j}{2} \tilde \psi^k \partial_k \right) \Big(\partial_i f_j - \partial_j f_i \Big) \right]\,.
\end{align}
For $\tdelta$ to be a symmetry, the last term of \C{Breaking-PS} proportional to $\fs_{ij}= \partial_i f_j - \partial_j f_i$, must vanish. In this case we obtain  the supercharge,
\begin{align} \label{tQ-class}
    \tilde Q = \psi_i\left(-\dot\phi^i - f^i\right),
\end{align} 
from (\ref{Breaking-PS}). 
This happens when $f_i = - \p_i V$ is a conservative force. 
We can explicitly check that
\begin{align} \label{Q-square-coord}
    & \tdelta_2 \tdelta_1 \phi_i = 0\,, && \tdelta_2 \tdelta_1 \psi_i = 0\,, && \tdelta_2 \tdelta_1 \tilde \psi_i = \tilde\epsilon_2\tilde \epsilon_1 \Big(\partial_j f_i - \partial_i f_j \Big) \psi^j\,.
\end{align}
So $\tdelta$ is nilpotent if $\mathcal{F}_{ij}=0$; for example in the case of a reciprocal system.
Furthermore using the equations of motion we see that
\begin{align}   \label{QtildeQ-Coordinates}
    &\begin{aligned} 
    & [\,\delta , \tdelta \,] \phi^i = \epsilon \tilde \epsilon \left( - 2 \dot \phi^i \right), \\
    &[\,\delta , \tdelta \,] \psi^i = \epsilon \tilde \epsilon \left(-2 \dot \psi^i - (\partial^i f^j - \partial^j f^i ) \psi_j \right), \\ & [\,\delta , \tdelta \,] \tilde \psi^i = \epsilon\tilde \epsilon \left( - 2 \dot{\tilde{\psi}}^i\right).    \end{aligned}
\end{align}
If $\fs_{ij} = 0$ then both $\delta$ and $\tdelta$ are symmetries and we can use them to define two supersymmetries which square to time translations. This is the structure of $N=2$ SUSY quantum mechanics found by Parisi and Sourlas.

To close this discussion we remark that for a reciprocal theory where $f^i =- \partial^i V$, the action (\ref{MSR-Action})  can be explicitly written in superspace with coordinates $(t, \theta, {\tilde \theta})$ using superfields, 
\begin{align} \label{22superfield}
    &\Phi^i = \phi^i + \psi^i \tilde \theta + \theta \tilde \psi^i + \theta \tilde \theta F^i\,.
\end{align}
The Grassmann coordinates satisfy $\theta^2 = (\tilde\theta)^2=0$. The superspace action takes the form, 
\begin{align} \label{S-MSR-Superspace}
    S_{MSR} & = \int dt d\tilde \theta d \theta \left( \frac{\tilde D \Phi^i D\Phi_i}{2} - V(\Phi) \right)\,, \\ \label{Ac-Rec-F}
    & = \int dt \left( - \frac{F_i F^i}{2} - F_i(\dot \phi^i + \partial^i V) + \tilde \psi^i \dot \psi_i - \tilde \psi^i \psi^j\, \partial_i \partial_j V \right)\,,
\end{align}
where the supercovariant derivatives and supercharges are
\begin{align}
\begin{aligned} \label{n=2susy}
    & D = \frac{\partial}{\partial \theta} + 2 \tilde \theta \frac{\partial}{\partial t}\,,\quad && \tilde D = \frac{\partial}{\partial \tilde \theta}\,, \\
    & Q = - \frac{\partial}{\partial \theta}\,,\quad && \tilde Q = -\frac{\partial}{\partial \tilde \theta} + 2 \theta \frac{\partial}{\partial t}\,.
\end{aligned}
\end{align}
The Hamiltonian generates the flow $- \frac{\p}{\p t}$ for stochastic time and from \C{n=2susy} we see that\footnote{In Euclidean signature, we Wick rotate time $t \rightarrow -i t$ so $H$ generates $- \frac{\p}{\p t}$.}
\begin{align}
    &\{Q,\tilde Q\} = -2 \frac{\p}{\p t} \,, \qquad \{D,\tilde{D}\} = 2 \frac{\p}{\p t}\,,
\end{align}
with all other anti-commutators vanishing. In writing the component fields of the superfield $\Phi^i$ in \C{22superfield}, we have followed standard notion in the supersymmetry literature. However, we note that the auxiliary field $F^i$ is essentially the response field ${\hat \phi}^i$ of \C{MSR-path-integral-response-field}.  

\subsection{A supersymmetric description of nonreciprocal theories} \label{Sec-SUSY-NR}

To show that nonreciprocal interactions are also supersymmetric, we return to \C{MSRpartition}. We will give an alternative formulation of this path-integral using superspace that has one real manifest supersymmetry.

To define our $N=1$ SUSY quantum mechanical theory in superspace we use the following superspace coordinates,
\begin{align}
    &\Lambda^i = \psi^i + \theta F^i\,, \qquad \varphi^i = \phi^i + \theta \chi^i\,.
\end{align}
Here $\phi^i$ is a real bosonic field while $(\psi^i, \chi^i)$ are real fermions. The SUSY covariant derivative and supercharge take the form,
\begin{align}\label{superspaceaction}
    &D_\chi = - \partial_\theta - \theta \partial_t\,, \qquad Q_\chi = -\partial_\theta + \theta \partial_t\,.
\end{align}
Notice that $Q_\chi^2 = - \partial_t$, $D_\chi^2  = \partial_t$ and $\{Q_\chi,D_\chi\} = 0$. The Lagrangian for the system is fermionic with the action in superspace given by
\begin{align} 
    \label{(0,1)Inspired-QM}
    S &= \int dt d\theta \left(\frac{1}{2} \Lambda^i D_\chi \Lambda_i - \Lambda_i E^i(\varphi) \right)\,, \\ \label{N=1-SUSY-QM}
    &=\int dt \left(-\frac{F_iF^i}{2} - F_i (\dot \phi^i - f^i) + \frac{1}{2} \psi_i \dot \psi^i + \psi^i ( \partial_t \delta_{ij} - \partial_j f_i) \chi^j \right)\,,
\end{align}
where $E^i(\phi) = \dot\phi^i - f^i$. This action is manifestly supersymmetric under transformations of $\varphi^i$ and $\Lambda^i$ induced by $Q_\chi$ via
\begin{align}
    \varphi^i \rightarrow \varphi^i + \com{\epsilon Q_\chi}{ \varphi^i}\,, \qquad \text{and} \qquad \Lambda^i \rightarrow \Lambda^i + \com{\epsilon Q_\chi}{ \Lambda^i}\,,
\end{align}
where $Q_\chi$ acts according to \C{superspaceaction}. This structure can be viewed as a generalization of (\ref{S-MSR-Superspace}) which does not require $f^i= - \partial^i V$ for the theory to possess at least one supercharge. On the fundamental fields, this supersymmetry is implemented by the transformations:
\begin{align} \label{Q-chi-coord}
    & \delta_\chi \phi^i = -\epsilon \chi^i\,, \qquad \delta_\chi \psi^i = -\epsilon F^i\,, \qquad \delta_\chi \chi^i = -\epsilon \dot \phi^i\,, \qquad \delta_\chi F^i = -\epsilon \dot \psi^i\,.
\end{align}
If we let $\epsilon$ be time-dependent to calculate the Noether charge, we find
\begin{align}
    \delta_\chi S & = \int dt \left[-\frac{d\epsilon}{dt} \big( -\psi_i f^i - \chi_i F^i \big) - \frac{d}{dt} \big(\epsilon \psi_i f^i\big) \right]\,,
\end{align}
which tells us that the Noether charge for this symmetry is $Q_{\chi} =  - \psi_i f^i - \chi_i F^i $. Therefore (\ref{N=1-SUSY-QM}) is a modified version of (\ref{Ac-Rec-F}) which does not possess  the BRST symmetry (\ref{Q-In-MSR}), which gives no dynamical information. Instead this action is symmetric under (\ref{Q-chi-coord}) which squares to the Hamiltonian and is therefore a  supersymmetry of the system.

\subsubsection*{\ul{\it Alternative representation of the functional determinant}}

At this stage it is not obvious that (\ref{N=1-SUSY-QM}) is related to the original stochastic theory (\ref{Langevin-General}). To prove that they are related, we need to show that the fermionic part of (\ref{N=1-SUSY-QM}) is just a rewriting of $\det(\frac{\delta E_a}{\delta \phi_b})$ in equation (\ref{MSR-path-integral}). This means revisiting the fermionic path integral formulation of the functional determinant, and it will be helpful to first start with the finite-dimensional case. 

Imagine we want to compute $\det(B)$ for a general finite-dimensional matrix $B$. We can express the determinant in terms of a Pfaffian using, 
\begin{align} \label{det(B)-standard}
    \text{Pf}\begin{pmatrix}
        0 & B \\
        -B^T & 0
    \end{pmatrix} = (-1)^{\frac{n(n-1)}{2}}\det(B) \,.
\end{align}
Is this a unique representation of the determinant in terms of a Pfaffian? We will argue there is a more general expression by first noting that 
\begin{align}
    \text{Pf}\begin{pmatrix}
        M & B \\
        -B^T & N
    \end{pmatrix} & = \text{Pf}(M) \text{Pf}(N+B^T M^{-1} B)\,,
\end{align}
where $M,N$ are skew-symmetric and $M$ is invertible, which requires that $M$ be even-dimensional.\footnote{In Appendix \ref{App-FermionicProofPffafian} we provide a more general proof of the identity (\ref{pfrelation}) using  Grassmann integration, which applies even when $M$ is not invertible, including when $M$ is odd-dimensional.} If we set $N=0$ and note that $\left[\text{Pf}(B^T M^{-1} B) \right]^2 = \det(B^T) \det(M^{-1}) \det(B)$ then $\text{Pf}(B^T M^{-1} B) = \det(B) \text{Pf}(M^{-1})$ which gives\footnote{Here we used that if $M$ is a $n\times n$ matrix with $n=2m$ then
\begin{align}
    \text{Pf}(M^{-1}) = \frac{(-1)^{\frac{n}{2}}}{\text{Pf}(M)} = \frac{(-1)^{ \frac{n(n-1)}{2} } }{\text{Pf}(M)}\,,
\end{align}
since $(-1)^{\frac{n}{2}} = (-1)^{\frac{n(n-1)}{2}}$ if $n$ is even. }
\begin{align} \label{pfrelation}
    \text{Pf}\begin{pmatrix}
        M & B \\
        -B^T & 0
    \end{pmatrix} = \text{Pf}(M) \text{Pf}(M^{-1}) \det(B) = (-1)^{\frac{n(n-1)}{2}} \det(B)\, .
\end{align}
So there is ambiguity in how we represent the determinant encoded in the choice of a skew-symmetric invertible matrix $M$. Using fermions, we can write 
\begin{align} \label{det(B)-New}
    \det(B)  = (-1)^{\frac{n(n-1)}{2}}\int  d\chi d\psi \, \exp \left[- \frac{1}{2} (\psi \ \ \chi )\begin{pmatrix}
       M & B \\
       -B^T & 0
    \end{pmatrix} \begin{pmatrix}
        \psi \\ \chi
    \end{pmatrix} \right] \,.
\end{align}

The proof outlined between equations (\ref{det(B)-standard}) and (\ref{det(B)-New}) and in Appendix \ref{App-FermionicProofPffafian} is valid for finite-dimensional systems. In Appendix \ref{timediscrete}, we extend the proof to the infinite-dimensional case. Absorbing any field-independent phase into the integration measure, we get
\begin{align} \label{New-Regularization}
    \det\left(\frac{\delta E^i}{\delta \phi_j}\right) & = \text{Pf} \begin{pmatrix}
       \partial_t \delta^{ij} & \partial_t\delta_{j}^i - \partial_j f^i \\
       \partial_t\delta_{i}^j + \partial_i f^j & 0
    \end{pmatrix}\,, \nonumber \\
    & = \int D \chi D\psi \exp \left[- \frac{1}{2} \int dt\, (\psi_i \ \ \chi^i )\begin{pmatrix}
       \partial_t \delta^{ij} & \partial_t\delta_{j}^i - \partial_j f^i \\
       \partial_t\delta_{i}^j + \partial_i f^j & 0
    \end{pmatrix} \begin{pmatrix}
        \psi_j \\ \chi^j
    \end{pmatrix} \right]\,, \nonumber \\
    & = \int D\chi D\psi \exp \left(- \int dt \left[ \frac{1}{2} \psi^i \dot \psi_i + \psi_i \dot \chi^i - \psi_i \partial_j f^i \chi^j \right]\right)\,.
\end{align}
Therefore (\ref{N=1-SUSY-QM}) is as valid a description of the stochastic system as (\ref{MSR-Action}). However we now have a supercharge $Q_\chi$ which squares to the Hamiltonian, even in the nonreciprocal case.

\subsubsection*{\ul{\it Integrating out the auxiliary field}}

After integrating out the auxiliary variable $F^i$ we obtain the $N=1$ SUSY action expressed in terms of $(V, A)$ of \C{defA},
\begin{align} \label{S-New-No-Auxiliary}
    S & = \int dt \left[ \frac{\dot \phi_i^2}{2} + \dot \phi^i A_i + \frac{(\partial_iV + A_i)^2}{2} +\frac{1}{2} \psi^i \dot \psi_i + \chi^i \dot \psi_i - \chi^j (\partial_j\partial^iV + \partial_j A^i)  \psi_i \right]\,.
\end{align}
It is interesting to note that the vector field $A_i$, which encodes the nonreciprocity, couples like an abelian gauge-field or Berry connection to the particle $\phi^i$ which carries an imaginary electric charge $e=i$.\footnote{Wick rotating the action for a charged particle from Minkowski space to Euclidean space, ${ i S^L = i\int dt \left(\frac{\dot\phi_i^2}{2} + e \dot \phi^i A_i \right) \xrightarrow[]{t\rightarrow -i t} - S^E = - \int dt \left(\frac{\dot\phi_i^2}{2} - i e \dot \phi^i A_i \right)}$, tells us that the electric charge $e=i$ in (\ref{MSR-Action-with-A}).}

The explicit form of the supersymmetry transformations (\ref{Q-chi-coord}) in these variables is given by,
\begin{align} \label{Q-chi-coord-AfterInt}
    & \delta_{\chi} \phi_i = - \epsilon \chi_i\,, \qquad \delta_{\chi} \psi_i = \epsilon \left(\dot \phi_i - f_i\right)\,, \qquad \delta_{\chi} \chi_i = -\epsilon \dot \phi_i\,.
\end{align}
The explicit form of the supercharge is,
\begin{align} \label{Qchi-Noether}
    &Q_\chi = -\chi_i ( i\pi^i + f^i ) - \psi_i f^i\,.
\end{align}
where again $\pi^i = i \dot \phi^i$ is the physical momentum. This is a non-Hermitian quantum mechanical system whose supersymmetry is implemented by a non-Hermitian supercharge. To close this section, we list the classical equations of motion of this system,
\begin{align}
\begin{aligned}
    &\ddot \phi^i = (\partial^i A_j - \partial_j A^i) \dot \phi^j + \partial^i \left( \frac{f_j^2}{2} \right) + \frac{\partial^i \partial_k f^ j}{2} [\chi^k,\psi_j]\,, \\
    &\dot\psi_i = -\partial_i f^j \psi_j\,,\\
    &\dot\chi^i = \partial_j f^i \chi^j + \partial^i f^j \psi_j\,.
\end{aligned}
\end{align}
At first glance, the role played by $\chi^i$ seems superficially similar to the role played by $\tpsi^i$ appearing in the standard MSR formalism; however, its equation of motion is quite different.

\subsection{Euclidean Quantization}

Now we want to describe the theory in the operator formalism. We start with the action \C{S-New-No-Auxiliary}, where a total derivative term has already been dropped, and we integrate in the canonical momentum:
\begin{align} \label{New-Action-with-momentum}
   & \int Dp_i D\phi_i D\psi_i D\chi_i  \, \times \cr &\exp\left( - \int dt \left[- i \dot \phi^i \, p_i + \frac{1}{2} \psi^i \dot \psi_i + \chi^i \dot \psi_i + \frac{1}{2}(p_i -iA_i)^2 + \frac{f_i^2}{2} + \chi^j \partial_j f^i \psi_i \right] \right)\,.
\end{align}
After symmetrizing the fermions the Hamiltonian of the system is,
\begin{align} \label{New-Hamiltonian}
     H = \frac{1}{2}(p_i -i A_i)^2 + \frac{1}{2}(\partial_i V + A_i )^2 - (\partial_j \partial^i V + \partial_j A^i)\frac{[ \chi^j , \psi_i ]}{2} \,,
\end{align}
and the canonical momenta are given by
\begin{align}
    p_i = i\frac{\partial L}{\partial \dot \phi^i} = i(\dot \phi_i + A_i)\,, \qquad p_i^\psi = i\frac{\partial L}{\partial \dot \psi^i} = - \frac{i}{2}(\psi_i + \chi_i )\,, \qquad p_i^\chi = i\frac{\partial L}{\partial \dot \chi^i} = - \frac{i}{2} \psi^i\,.
\end{align}
For convenience we also define the physical momentum $\pi_i = i \dot \phi_i = p_i - i A_i$. The quantization of the bosonic degrees of freedom is done by imposing $[\phi^i,p_j] = i \delta^i_j$ with $\phi^i$ and $p_i$ commuting with all other variables. 

The fermion quantization conditions are,
\begin{align} \label{Standard-Comm-Rel}
    &\{\psi_i , p_j^\psi \} = - \frac{i}{2} \delta_{ij}\,,
    \qquad \{\chi_i , p_j^\psi \} = 0\,, \qquad \{\chi_i , p_j^\chi \} = - \frac{i}{2} \delta_{ij}\,,
    \qquad \{\psi_i , p_j^\chi \} = 0\,.
\end{align}
In terms of $\psi_i$ and $\chi^i$ and the bosonic momenta the commutation relations are,
\begin{align} \label{Bosonic-Commutation}
    &[\phi^i,\pi_j] = i\delta^i_{j}\,, \qquad [\pi_i,\pi_j] = \mathcal{F}_{ij}\,, \\ \label{Fermionic-Commutation}
    &\{ \chi^i, \psi_j \} = \delta^i_{j}\,, \qquad \{\psi_i,\psi_j \} = 0\,, \qquad \{\chi^i, \chi^j \} = -\delta^{ij}\,.
\end{align}

\subsubsection*{\ul{\it Comparison with the MSR Hamiltonian}}

If we perform canonical quantization of the MSR action (\ref{MSR-Action}) we find the Hamiltonian,
\begin{align}
     H_{MSR} = \frac{1}{2}(p_i -i A_i)^2 + \frac{1}{2}(\partial_i V + A_i )^2 - (\partial_j \partial^i V + \partial_j A^i)\frac{[ \tpsi^j , \psi_i ]}{2} \,.
\end{align}
For MSR the bosonic degrees of freedom satisfy the same commutation relations (\ref{Bosonic-Commutation}). Meanwhile the fermions $\tpsi^i$ and $\psi_i$ satisfy
\begin{align} \label{Comm-Rel-Fermions-MSR}
    \{\tpsi_i,\psi_j\} = \delta_{ij}\,, \qquad  \{\psi_i,\psi_j\} =  \{\tpsi_i,\tpsi_j\} = 0\,.
\end{align}
Even though the Hamiltonians look similar, the fermion structure is substantially different. Note that the combination $\tilde \psi_i - \hlf \psi_i$ satisfies the following relation, 
\begin{align} \label{chi-chi-and-tpsi}
    \left\{\tilde \psi_i - \frac{\psi_i}{2}, \tilde \psi_j - \frac{\psi_j}{2} \right\} = - \{\tilde \psi_i , \psi_j \} = - \delta_{ij}\,.
\end{align}
Therefore the fermion $\chi_i$, which appears in our new formalism, can be written in terms of MSR fermions as
\begin{align}
    \chi_i = \tilde \psi_i - \frac{\psi_i}{2}\,.
\end{align}
Using the relations, 
\begin{align}
    [\chi^j,\psi_i] = [\tilde \psi^j , \psi_i] -\frac{1}{2}[\psi^j,\psi_i] = [\tilde \psi^j , \psi_i] - \psi_j \psi_i\,,
\end{align}
and $(\partial_i \partial_j V) \psi_j \psi_i = 0$, we can rewrite the Hamiltonian of \C{New-Hamiltonian} in terms of MSR fermions:
\begin{align} \label{NR-Hamiltonian}
    H = \frac{(p_i -i A_i)^2}{2} + \frac{(\partial_i V + A_i )^2}{2} - \frac{\partial_j \partial^i V + \partial_j A^i}{2} [ \tilde \psi^j , \psi_i ] + \frac{\partial_j A_i-\partial_i A_j}{4} \psi_j \psi_i\,.
\end{align}
The last term of \C{NR-Hamiltonian} is the novelty with respect to the usual MSR Hamiltonian and it requires nonreciprocity. This coupling does not preserve fermion number, which is conserved by the MSR Hamiltonian.  

The BRST charge $Q$ and the fermionic operator $\tilde Q$ obtained from (\ref{Q-class}) and(\ref{tQ-class}), respectively, are given by:
\begin{align} \label{Q-and-tQ-MSR}
    Q = \tpsi^i \left( -i\pi_i - f_i \right)\,, \qquad \tilde Q = \psi^i \left( i\pi_i - f_i \right)\,.
\end{align}
It is easy to check that
\begin{align} \label{Qsquare-tQsquare-and-H}
    Q^2 = 0\,, \qquad \tilde Q^2 = - \mathcal{F}_{ji} \psi_j \psi_i\,, \qquad \{Q,\tilde Q\} = 2 H_{MSR}\,.
\end{align}
Even though $Q$ and $\tilde Q$ anticommute to $H_{MSR}$, the operator $\tilde Q$ does not square to zero and does not commute with $H_{MSR}$. This is why the MSR description of a nonreciprocal theory is not supersymmetric as we discussed in Section \ref{Sec-magic-of-reciprocity}.\footnote{ \label{subtle} A subtle point is that even though $\{Q,\tilde Q\} = 2 H_{MSR}$, $[\delta,\tdelta] \neq -2\epsilon \tilde\epsilon\partial_t$ as we saw in (\ref{QtildeQ-Coordinates}). This happens because $\tdelta\mathcal{O}\neq[\tepsilon \tilde Q,\mathcal{O}]$; that is $\tilde Q$ does not implement $\tdelta$. As explained in Appendix \ref{symplectic}, the vector field $\tdelta$ is not a Hamiltonian vector field. There is therefore no scalar operator that can implement it via commutation.}

Using the definition of $\chi^i$ in terms of $\psi_i$ and $\tpsi^i$ we can write the supercharge (\ref{Qchi-Noether}) as
\begin{align}
    Q_\chi & = Q + \frac{\tilde Q}{2}\,,
\end{align}
with $Q$ and $\tilde Q$ given by (\ref{Q-and-tQ-MSR}). It is easy to check that
\begin{align}
    Q_\chi^2
    & = \frac{\{ Q ,\tilde Q \}}{2} + \frac{\tilde Q^2}{4} = H_{MSR} - \frac{\mathcal{F}_{ji}}{4}\psi_j \psi_i \,.
\end{align}
Finally we express our supersymmetric action in terms of MSR fermions, 
\begin{align} \label{New-SUSy-S-new-coord}
    S = \int dt \left(\frac{(\dot \phi_i - f_i)^2}{2} + \tilde \psi^i \dot \psi_i +\tilde \psi^j \partial_j f^i \psi_i - \frac{\partial^j f^i}{2} \psi_j \psi_i \right)\,.
\end{align}

\section{Ward Identities and the Fluctuation-Dissipation Theorem} \label{Sec-Ward}

The supersymmetries and the BRST symmetry identified in Section \ref{MSR} are not merely formal properties of the action. Like any symmetry of a path integral, they constrain the correlation functions of the theory through Ward identities: exact, non-perturbative relations among the correlation and response functions of the stochastic system. These identities are the natural place to look for physical, observable consequences of the supersymmetric structure. In the reciprocal case the Ward identity of the BRST charge $Q$ enforces causality and the conservation of probability, while the Ward identity of the second supercharge $\tilde Q$ is precisely the fluctuation-dissipation theorem. When the interactions are nonreciprocal, $\tilde Q$ ceases to be a symmetry and the fluctuation-dissipation theorem is modified by a term controlled by the non-reciprocity $\fs_{ij}$. This violation is a sharp, measurable signature of the departure from equilibrium.

In this section we develop the Ward identities of the theory. We first establish the general machinery, valid for any symmetry of the MSR action, and then reinstate the temperature $T$ in both the Parisi-Sourlas and the $N=1$ supersymmetric formulations, since the fluctuation-dissipation relations are conventionally stated with the temperature displayed.

Let $\Phi = (\phi, \hat \phi, \psi,\tilde \psi)$ and consider a general transformation of the fields $\Phi \rightarrow \Phi + \delta \Phi$ in the path integral. Correlation functions are computed from
\begin{align} \label{Ward-Z}
    \langle \mathcal O\rangle = \int D\Phi\;\mathcal O[\Phi]\,e^{-S_{\text{MSR}}[\Phi]}\,,
\end{align}
normalized so that $\langle 1\rangle = 1$. Since $\Phi$ is only a dummy variable of integration, relabeling $\Phi \to \Phi + \delta\Phi$ cannot change $\langle\mathcal O\rangle$. Equating the integral before and after the relabeling, using $\delta\big(\mathcal O\,e^{-S_{\text{MSR}}}\big) = \big(\delta\mathcal O - \mathcal O\,\delta S_{\text{MSR}}\big)e^{-S_{\text{MSR}}}$ and $D(\Phi+\delta\Phi) = D\Phi\,\big(1 + \mathrm{Str}\,\tfrac{\p\,\delta\Phi}{\p\Phi}\big)$ for the Jacobian gives\footnote{Let $\mathrm{Sdet}\, M$ and $\mathrm{Str}\, M$ be the superdeterminant and supertrace of a matrix $M$. Standard matrix identities give $\mathrm{Sdet}(1+M) = e^{\mathrm{Str}\ln(1+M)} = 1 + \mathrm{Str}\,M + O(M^2)$, which is the first-order form quoted here. Recall that a Grassmann integration variable transforms with the inverse of its Jacobian factor, so the fermionic diagonal terms enter with the opposite sign, $\mathrm{Str}\,M = \sum_A (-1)^{|A|}\,\p\,\delta\Phi^A/\p\Phi^A$ with $|A|=0$ for $(\phi,\hat\phi)$ and $|A|=1$ for $(\psi,\tpsi)$.}
\begin{align} \label{Ward-Master}
    0 = \Big\langle\, \mathrm{Str}\,\frac{\p\,\delta\Phi}{\p\Phi} \;+\; \delta\mathcal O \;-\; \mathcal O\,\delta S_{\text{MSR}} \,\Big\rangle\,.
\end{align}
The first term is the graded Jacobian of the measure under the transformation. If the measure is invariant under this transformation (no anomalies), then for any observable $\mathcal O$ we have
\begin{align} \label{Ward-Identity}
    \langle \delta\,\mathcal O\rangle = \langle \mathcal O\,\delta S_{\text{MSR}}\rangle\,.
\end{align}
For the fermionic transformations used below, the anomaly-free condition is immediate in the variables used here. Each transformation maps a field into a different field, so no field appears in its own variation and every diagonal entry of the Jacobian vanishes. For example, for the Parisi-Sourlas variation $\delta\phi^i\propto\tpsi^i$, $\delta\psi^i\propto\dot\phi^i-f^i$, and $\delta\hat\phi^i=\delta\tpsi^i=0$, so
\begin{align} \label{No-Anomaly}
    \mathrm{Str}\,\frac{\p\,\delta\Phi}{\p\Phi} =
    \frac{\p(\delta\phi^i)}{\p\phi^i}
    + \frac{\p(\delta\hat\phi^i)}{\p\hat\phi^i}
    - \frac{\p(\delta\psi^i)}{\p\psi^i}
    - \frac{\p(\delta\tpsi^i)}{\p\tpsi^i}
    = 0\,.
\end{align}
The measure is invariant and there is no anomaly, so \C{Ward-Identity} holds exactly. If $\delta$ is a symmetry of the action, $\delta S_{\text{MSR}} = 0$, then we have the Ward identity. The Ward identities are exact, non-perturbative relations between the correlation and response functions of the stochastic system.

For a continuous symmetry the identity \C{Ward-Identity} takes the form of a local statement built from a conserved current. Let $\delta_\theta\Phi = i\,q \,\theta\,\Phi$ be a symmetry of the action and promote the parameter to a function of time, $\theta\to\theta(t)$. Since the transformation is a symmetry for constant $\theta$, the action can change only through derivatives of the parameter, and its variation defines a Noether current $j(t)$,
\begin{align} \label{Noether-Current}
    \delta_\theta S_{\text{MSR}} = \int dt\;\dot\theta(t)\,j(t)\,.
\end{align}
Consider an operator $\mathcal O = \prod_a \Phi_a(t_a)$ assembled from fields $\Phi_a$ inserted at times $t_a$, and let each field transform with a charge $q_a$ under the symmetry, $\delta_\theta\Phi_a(t_a) = i\,q_a\,\theta(t_a)\,\Phi_a(t_a)$. Using the Leibniz rule, and writing $\theta(t_a) = \int dt\,\delta(t-t_a)\,\theta(t)$, the left-hand side of \C{Ward-Identity} is
\begin{align} \label{Ward-LHS}
    \langle\delta_\theta\mathcal O\rangle = i\sum_a q_a\,\theta(t_a)\,\langle\mathcal O\rangle = i\int dt\;\theta(t)\,\sum_a q_a\,\delta(t-t_a)\,\langle\mathcal O\rangle\,.
\end{align}
The $\delta$-functions just record the time at which each charged operator sits. The right-hand side follows from \C{Noether-Current} after an integration by parts in $t$,
\begin{align} \label{Ward-RHS}
    \langle\mathcal O\,\delta_\theta S_{\text{MSR}}\rangle = \int dt\,\dot\theta(t)\,\langle j(t)\,\mathcal O\rangle = -\int dt\,\theta(t)\,\p_t\langle j(t)\,\mathcal O\rangle\,.
\end{align}
Since $\theta(t)$ is arbitrary, equating \C{Ward-LHS} and \C{Ward-RHS} gives the local Ward identity
\begin{align} \label{Ward-Current-Local}
    \p_t\big\langle j(t)\,\mathcal O\big\rangle = - i\sum_a q_a\,\delta(t-t_a)\,\langle\mathcal O\rangle \,.
\end{align}
The current is conserved inside correlators away from the insertions, where its divergence is sourced by the charge $q_a$ sitting there. Integrating over all $t$, the left-hand side is a total derivative with vanishing boundary terms, leaving the global selection rule
\begin{align} \label{Ward-Global}
    i\Big(\sum_a q_a\Big)\,\langle\mathcal O\rangle = 0\,,
\end{align}
that is, a correlator vanishes unless its charges balance out.

\subsection{Reinstating temperature} \label{Sec-temperature}

Throughout Section \ref{MSR} we set the noise amplitude  in \C{noisecorrel} to $\sigma = 1$. The fluctuation-dissipation relations, however, are conventionally quoted with the temperature displayed, so it is useful to reinstate it. We work at a general temperature $T$, for which the noise correlator is
\begin{align} \label{noise-T}
    \langle\xi_i(t)\,\xi_j(t')\rangle = 2T\,\delta_{ij}\,\delta(t-t')\,,
\end{align}
so that the conventions of Section \ref{MSR} correspond to $T = \tfrac12$. The Gaussian noise is then drawn from $P[\xi]\propto\exp\!\big(-\tfrac{1}{4T}\int dt\,\xi_i\xi^i\big)$.

\subsubsection*{\ul{\it The Parisi-Sourlas construction}}

Repeating the MSR construction of Section \ref{MSR} with the noise \C{noise-T}, and using the response field $\hat\phi^i$ to implement the Langevin constraint, gives  equivalent forms for the action
\begin{align} \label{PS-action-aux-T}
    S_{\text{MSR}} &= \int dt \left[ T\,\hat\phi_i\hat\phi^i - i\,\hat\phi_i\big(\dot\phi^i - f^i\big) + \tpsi^i\dot\psi_i + \tpsi^j\,\p_j f^i\,\psi_i \right]\,, \\ \label{PS-action-T}
    S_{\text{MSR}} &= \int dt \left[ \frac{(\dot\phi^i)^2}{4T} + \frac{\dot\phi^i A_i}{2T} + \frac{(f^i)^2}{4T} + \tpsi^i \dot\psi_i + \tpsi^j \p_j f^i\, \psi_i \right]\,,
\end{align}
with the second following from the first after integrating out $\hat\phi^i$ through its algebraic equation of motion $\hat\phi^i = \tfrac{i}{2T}(\dot\phi^i - f^i)$ and ignoring the total derivative. At $T = \tfrac12$ these are the MSR actions \C{MSR-path-integral-response-field} and \C{MSR-Action}. The transformations \C{Q-In-MSR} and \C{Qtilde-In-MSR} become
\begin{align} \label{PS-delta-T}
    &\delta\phi^i = -\epsilon\,\tpsi^i\,, && \delta\psi^i = \frac{1}{2T}\,\epsilon\big(\dot\phi^i - f^i\big)\,, && \delta\tpsi^i = 0\,, \\ \label{PS-tdelta-T}
    &\tdelta\phi^i = \tepsilon\,\psi^i\,, && \tdelta\psi^i = 0\,, && \tdelta\tpsi^i = \frac{1}{2T}\,\tepsilon\big(-\dot\phi^i - f^i\big)\,.
\end{align}
The first is a symmetry for any force. The second is not: acting on the action it gives
\begin{align} \label{PS-obstruction-T}
    \tdelta S_{\mathrm{MSR}}
    = \int dt\,\tepsilon\left[
        \frac{1}{2T}\left(\dot{\phi}^i-f^i\right)\psi^j
        -\frac{\psi^i\psi^j}{2}\,\tpsi^k\partial_k
    \right]\mathcal{F}_{ij}
    +(\text{total derivative})\,.
\end{align}
This is the finite-temperature version of the obstruction \C{Breaking-PS}: the $(\dot\phi-f)\psi$ part carries the factor $1/(2T)$, while the ghost-curvature derivative term keeps its coefficient. The obstruction vanishes only in the reciprocal case $\fs_{ij}=0$. Canonical quantization gives the Hamiltonian
\begin{align} \label{PS-H-T}
    H_{MSR} = T\Big(p_i - \frac{i}{2T}A_i\Big)^2 + \frac{f_i f^i}{4T} + \frac{\p_j f^i}{2}\,[\tpsi^j,\psi_i]\,,
\end{align}
and the supercharges
\begin{align} \label{PS-QtQ-T}
    Q = \tpsi^i\Big(-i\pi_i - \frac{1}{2T}f_i\Big)\,, \qquad \tilde Q = \psi^i\Big(i\pi_i - \frac{1}{2T}f_i\Big)\,, \qquad \pi_i = p_i - \frac{i}{2T}A_i\,,
\end{align}
which satisfy
\begin{align} \label{PS-algebra-T}
    Q^2 = 0\,, \qquad \tilde Q^2 = -\frac{\fs_{ij}}{2T}\,\psi^i\psi^j\,, \qquad \{Q,\tilde Q\} = \frac{1}{T}\,H_{MSR}\,.
\end{align}
At $T = \tfrac12$ these reproduce \C{Q-and-tQ-MSR} and \C{Qsquare-tQsquare-and-H}.

\subsubsection*{\ul{\it The $N=1$ supersymmetric construction}}

The single supercharge of Section \ref{Sec-SUSY-NR} was built at $T=\tfrac12$ as $Q_\chi = Q + \tfrac12\tilde Q$, equivalently through the fermion $\chi_i = \tpsi_i - \tfrac12\psi_i$ satisfying \C{chi-chi-and-tpsi}. At a general temperature we write $Q_\chi = Q + c\,\tilde Q$, equivalently
\begin{align} \label{T-frame}
    \chi_i = \tpsi_i - c\,\psi_i\,, \qquad \{\chi^i,\chi^j\} = -2c\,\delta^{ij}\,.
\end{align}
The coefficient $c$ is fixed by demanding that $\delta_\chi$ square to a time translation, $\delta_\chi^2\phi^i = \epsilon_1\epsilon_2\,\dot\phi^i$, which is the defining property $Q_\chi^2 = H$ of the $N=1$ supersymmetry. Using \C{PS-delta-T} and \C{PS-tdelta-T},
\begin{align} \label{T-second-variation}
    \delta_\chi\psi^i = \frac{1}{2T}\,\epsilon\,(\dot\phi^i - f^i)\,, \qquad \delta_\chi\tpsi^i = \frac{c}{2T}\,\epsilon\,(-\dot\phi^i - f^i)\,,
\end{align}
so that $\delta_\chi^2\phi^i = -\epsilon_1\big(\delta_\chi\tpsi^i - c\,\delta_\chi\psi^i\big) = \tfrac{c}{T}\,\epsilon_1\epsilon_2\,\dot\phi^i$. Demanding $\delta_\chi^2\phi^i = \epsilon_1\epsilon_2\,\dot\phi^i$ forces $c = T$. The algebra and supercharge at temperature $T$ are therefore
\begin{align} \label{T-result}
    \chi_i = \tpsi_i - T\,\psi_i\,, \qquad \{\chi^i,\chi^j\} = -2T\,\delta^{ij}\,, \qquad Q_\chi = Q + T\,\tilde Q\,,
\end{align}
and squaring the supercharge gives
\begin{align} \label{T-Hsusy}
    Q_\chi^2 = H_{MSR} - \frac{T}{2}\,\fs_{ji}\,\psi_j\psi_i\,,
\end{align}
which reduces to $Q_\chi^2 = H_{MSR} - \tfrac14\fs_{ji}\psi_j\psi_i$ of Section \ref{Sec-SUSY-NR} at $T=\tfrac12$. In the $(\psi,\tpsi)$ frame the action at temperature $T$ reads
\begin{align} \label{SUSY-action-T}
    S = \int dt \left[ \frac{(\dot\phi^i)^2}{4T} + \frac{\dot\phi^i A_i}{2T} + \frac{(f^i)^2}{4T} + \tpsi^i\dot\psi_i + \tpsi^j\,\p_j f^i\,\psi_i - T\,\p^j f^i\,\psi_j\psi_i \right]\,,
\end{align}
invariant under $\delta_\chi = \delta + T\,\tdelta$, which acts as
\begin{align} \label{SUSY-delta-T}
    &\delta_\chi\phi^i = -\epsilon\big(\tpsi^i - T\psi^i\big)\,, && \delta_\chi\psi^i = \frac{\epsilon}{2T}\big(\dot\phi^i - f^i\big)\,, && \delta_\chi\tpsi^i = -\frac{\epsilon}{2}\big(\dot\phi^i + f^i\big)\,.
\end{align}
At $T = \tfrac12$ this is the $N=1$ action \C{New-SUSy-S-new-coord} and the supersymmetry \C{Q-chi-coord-AfterInt} of Section \ref{Sec-SUSY-NR}.

\subsection{Ward identities in the Parisi-Sourlas formulation} \label{Sec-PS-ward}

We now apply the machinery above to the MSR action, using the two Parisi-Sourlas transformations $\delta$ and $\tdelta$ of Section \ref{Sec-magic-of-reciprocity}. Two objects organize the discussion: the correlation function and the linear response function,
\begin{align} \label{CR-def}
    C^{ij}(t,t') = \langle\phi^i(t)\,\phi^j(t')\rangle\,, \qquad R^{ij}(t,t') = \frac{\delta\langle\phi^i(t)\rangle}{\delta h_j(t')}\bigg|_{h=0}\,,
\end{align}
where $R$ is the response to an infinitesimal probe force $f^i \to f^i + h^i$. Because $h^i$ enters the Langevin equation \C{Langevin-General} in the same way as the noise, the response is the field-noise cross-correlation
\begin{align} \label{Resp-Noise}
    R^{ij}(t,t') = \frac{1}{2T}\,\langle\phi^i(t)\,\xi^j(t')\rangle = \frac{1}{2T}\,\langle\phi^i(t)\,(\dot\phi^j - f^j)(t')\rangle = -i\,\langle\phi^i(t)\,\hat\phi^j(t')\rangle\,,
\end{align}
using the on-shell value $\hat\phi^j = \tfrac{i}{2T}(\dot\phi^j - f^j)$ of the response field.

\subsubsection*{\ul{\it Causality and probability conservation from the BRST charge $Q$}}

The BRST symmetry \C{PS-delta-T} holds for any force, so the measure is invariant and $\langle\delta\mathcal O\rangle = 0$ for every $\mathcal O$. Taking $\mathcal O = \phi^i(t)\,\psi^j(t')$,
\begin{align} \label{PS-brst-ward}
    0 = \big\langle \delta[\phi^i(t)\,\psi^j(t')]\big\rangle = -\epsilon\,\langle \tpsi^i(t)\,\psi^j(t')\rangle + \frac{\epsilon}{2T}\big\langle\phi^i(t)\,(\dot\phi^j - f^j)(t')\big\rangle\,.
\end{align}
The second term is the response \C{Resp-Noise}, so the mixed ghost two-point function is exactly the response,
\begin{align} \label{Resp-Ghost}
    R^{ij}(t,t') = -i\,\langle \phi^i(t)\,\hat\phi^j(t')\rangle = \langle \tpsi^i(t)\,\psi^j(t')\rangle\,.
\end{align}
The ghost propagator is retarded, so it vanishes for $t < t'$, which is the statement of causality,
\begin{align} \label{Causality}
    R^{ij}(t,t') = 0 \qquad (t < t')\,.
\end{align}
The response cannot precede the perturbation. Causality is protected by $Q$ and survives arbitrarily far from equilibrium.

The same charge enforces the conservation of probability. The response field is both BRST-exact and BRST-closed: from \C{PS-delta-T} it is the variation of a ghost, $\delta\psi^i = -i\epsilon\,\hat\phi^i$, while $\delta\hat\phi^i = 0$. Consider a string of response fields inserted at arbitrary times. Because every $\hat\phi$ is closed, the BRST variation of $\mathcal O = \psi^{i_1}(t_1)\,\hat\phi^{i_2}(t_2)\cdots\hat\phi^{i_n}(t_n)$ acts only on the lone ghost $\psi^{i_1}$,
\begin{align} \label{BRST-exact-string}
    \delta\mathcal O
    = \delta\psi^{i_1}(t_1)\,\hat\phi^{i_2}(t_2)\cdots\hat\phi^{i_n}(t_n)
    = -i\epsilon\,\hat\phi^{i_1}(t_1)\,\hat\phi^{i_2}(t_2)\cdots\hat\phi^{i_n}(t_n)\,.
\end{align}
Thus any product of response fields is itself BRST-exact. Its expectation value is the average of a BRST variation, which vanishes by the Ward identity $\langle\delta\mathcal O\rangle = 0$:
\begin{align} \label{Prob-Cons}
    \big\langle\hat\phi^{i_1}(t_1)\cdots\hat\phi^{i_n}(t_n)\big\rangle = 0\,, \qquad n\geq 1\,.
\end{align}
This holds for any force, reciprocal or not: no correlator built purely from response fields can be nonzero.

The physical content of \C{Prob-Cons} is the conservation of probability. Couple a probe force $h_i(t)$ to the response field and form the generating functional
\begin{align} \label{Z-probe}
    Z[h] = \big\langle e^{-i\int dt\,h_i(t)\,\hat\phi^i(t)}\big\rangle\,.
\end{align}
In the MSR action \C{MSR-path-integral-response-field}, the response field couples to the dynamics through $-i\hat\phi_i(\dot\phi^i-f^i)$, so the insertion in \C{Z-probe} shifts the drift $f^i\to f^i+h^i$. Therefore $Z[h]$ is the normalization of the path integral, namely the total probability, in the presence of the applied force $h^i$. Expanding in powers of $h$,
\begin{align} \label{Z-taylor}
    Z[h] = \sum_{n=0}^\infty \frac{(-i)^n}{n!}
    \int dt_1\cdots dt_n\,
    h^{i_1}(t_1)\cdots h^{i_n}(t_n)\,
    \big\langle\hat\phi_{i_1}(t_1)\cdots\hat\phi_{i_n}(t_n)\big\rangle\,,
\end{align}
every term with $n\geq1$ vanishes by \C{Prob-Cons}. Only the $n=0$ term survives,
\begin{align} \label{Z-conserved}
    Z[h] = Z = \langle 1\rangle = 1\,,
\end{align}
independent of $h$. An applied force can redistribute probability among configurations but cannot change the total probability. This probability-preserving property is guaranteed nonperturbatively by the BRST charge $Q$.

\subsubsection*{\ul{\it Fermion number and the $U(1)$ symmetry}}

Both fermion terms of the action \C{PS-action-T} carry one $\psi$ and one $\tpsi$, so for any force the action is invariant under the phase rotation $(\psi^i,\tpsi^i) \to (e^{i\theta}\psi^i, e^{-i\theta}\tpsi^i)$, whose infinitesimal form is
\begin{align} \label{U1-transformation}
    \delta_\theta\psi^i = i\theta\,\psi^i\,, \qquad
    \delta_\theta\tpsi^i = -i\theta\,\tpsi^i\,, \qquad
    \delta_\theta\phi^i = \delta_\theta\hat\phi^i = 0\,.
\end{align}
Promoting $\theta\to\theta(t)$, only the kinetic term produces a derivative of the parameter, isolating the conserved fermion-number current and charge
\begin{align} \label{Fermion-Number}
    \delta_\theta S_{\text{MSR}} = \int dt\,\dot\theta(t)\,j(t)\,, \qquad j = i\,N_F\,, \qquad N_F = \tpsi^i\psi_i\,.
\end{align}
Substituting the charges $q_a = +1$ for each $\psi$, $-1$ for each $\tpsi$, and $0$ for each boson into the global identity \C{Ward-Global} gives the ghost-number selection rule
\begin{align} \label{Selection-PS}
    \big\langle \psi^{i_1}\cdots\psi^{i_{n_\psi}}\,\tpsi^{j_1}\cdots\tpsi^{j_{n_{\tpsi}}}\,(\text{bosons})\big\rangle = 0 \qquad\text{unless}\qquad n_\psi = n_{\tpsi}\,.
\end{align}
Only ghost-number-neutral correlators survive; in particular $\langle\psi^i\psi^j\rangle = \langle\tpsi^i\tpsi^j\rangle = 0$, while the neutral bilinear $\langle\tpsi^i\psi^j\rangle = R^{ij}$ is the response \C{Resp-Ghost}.

\subsubsection*{\ul{\it The fluctuation-dissipation theorem and its violation}}

The second transformation $\tdelta$ \C{PS-tdelta-T} is a symmetry only in the reciprocal case; out of equilibrium it leaves the obstruction \C{Breaking-PS}, so its Ward identity carries a source. Taking $\mathcal O = \phi^i(t)\,\tpsi^j(t')$ in \C{Ward-Identity},
\begin{align} \label{FDT-ward}
    \langle\psi^i(t)\,\tpsi^j(t')\rangle - \frac{1}{2T}\big\langle\phi^i(t)\,(\dot\phi^j + f^j)(t')\big\rangle = \Sigma^{ij}(t,t')\,,
\end{align}
where $\Sigma^{ij} = \tfrac{1}{\tepsilon}\langle\phi^i(t)\,\tpsi^j(t')\,\tdelta S_{\text{MSR}}\rangle$ is proportional to $\fs_{ij}$ by \C{Breaking-PS} and vanishes for a reciprocal force. Writing $\dot\phi^j + f^j = 2\dot\phi^j - \xi^j$, using $\langle\phi^i\xi^j\rangle = 2T\,R^{ij}$ and $\langle\psi^i(t)\tpsi^j(t')\rangle = -R^{ji}(t',t)$, this becomes the exact relation
\begin{align} \label{FDT-alltime}
    R^{ij}(t,t') - R^{ji}(t',t) - \frac{1}{T}\,\p_{t'}C^{ij}(t,t') = \Sigma^{ij}(t,t')\,.
\end{align}
Causality \C{Causality} removes $R^{ji}(t',t)$ for $t>t'$; in the stationary state, with $\tau = t - t' > 0$,
\begin{align} \label{FDT-stationary}
    R(\tau) = -\frac{1}{T}\,\p_\tau C(\tau) + \Sigma(\tau)\,, \qquad \tau > 0\,.
\end{align}
The linear response is fixed by the spontaneous fluctuations up to the term $\Sigma$ generated by insertions of $\fs$: a nonreciprocal interaction can therefore show up as a measurable breakdown of the fluctuation-dissipation theorem. In the reciprocal case $\fs_{ij}=0$, the transformation $\tdelta$ is a genuine supersymmetry, $\Sigma=0$, and \C{FDT-alltime} reduces to the equilibrium fluctuation-dissipation theorem. The converse need not hold component by component: for nonzero $\fs$, symmetries, degeneracies, special observables, or cancellations in the trajectory average can still make a particular $\Sigma^{ij}(t,t')$ vanish. If $\Sigma =0$ then we recover exact fluctuation-dissipation relations:
\begin{align} \label{FDT-rec}
    R^{ij}(t,t') - R^{ji}(t',t) = \frac{1}{T}\,\p_{t'}C^{ij}(t,t')\,, \qquad R(\tau) = -\frac{1}{T}\,\p_\tau C(\tau)\quad(\tau>0)\,.
\end{align}
The same procedure applied to $\mathcal O = \phi^{i_1}(t_1)\cdots\phi^{i_n}(t_n)\,\tpsi^j(t')$ generates the higher-point fluctuation-dissipation relations, each an exact identity relating the response of an $n$-point function to the time derivative of the $(n{+}1)$-point correlator, up to a violation $\Sigma_{n+1}\propto\fs$.

\subsection{Ward identities in the \texorpdfstring{$N=1$}{N=1} supersymmetric formulation} \label{Sec-N1-ward}

In the $N=1$ formulation the single supercharge $Q_\chi$ is a symmetry of $S_{\text{SUSY}}$ for every force, reciprocal or not. Hence $\delta_\chi S_{\text{SUSY}} = 0$ and, by \C{Ward-Identity},
\begin{align} \label{Ward-NoObstruction}
    \langle \delta_\chi\mathcal O\rangle = 0 \qquad\text{for every observable } \mathcal O\,.
\end{align}
None of its Ward identities carries a source: each is an exact relation with a vanishing right-hand side, arbitrarily far from equilibrium. This is the structural contrast with the Parisi-Sourlas analysis, where $\tdelta$ is a symmetry only at $\fs_{ij}=0$.

\subsubsection*{\ul{\it A fermion-number selection rule}}

In the $(\psi,\tpsi)$ frame the two actions differ by a single ghost bilinear, from \C{SUSY-action-T} and \C{PS-action-T},
\begin{align} \label{SUSY-PS-Split}
    S_{\text{SUSY}} = S_{\text{MSR}} - \frac{T}{2}\int dt\;\fs_{ji}\,\psi^j\psi^i\,.
\end{align}
The vertex $\frac{T}{2}\fs_{ji}\psi^j\psi^i$ raises the fermion number $N_F$ \C{Fermion-Number} by two units. The supersymmetric average is the Parisi-Sourlas average dressed by this vertex,
\begin{align} \label{SUSY-Average-Expansion}
\begin{aligned}
    \langle X\rangle_{\text{SUSY}}
    &= \Big\langle X\,\exp\left(\frac{T}{2}\int dt\,\fs_{ji}\,\psi^j\psi^i\right)\Big\rangle_{\text{MSR}} \\
    &= \sum_{m=0}^{\infty}\frac{1}{m!}\left(\frac{T}{2}\right)^m
       \Big\langle X\,\left(\int dt\,\fs_{ji}\,\psi^j\psi^i\right)^m\Big\rangle_{\text{MSR}}\,.
\end{aligned}
\end{align}
Each vertex supplies two $\psi$'s. By the Parisi-Sourlas selection rule \C{Selection-PS}, the $m$-vertex term survives only if its total ghost number vanishes, equivalently if $\#\tpsi(X)-\#\psi(X)=2m$. Hence $\langle X\rangle_{\text{SUSY}}$ is non-zero only when
\begin{align} \label{Selection-N1}
    \#\tpsi(X) - \#\psi(X) = 2m\,, \qquad m = 0,1,2,\dots\,,
\end{align}
which is strictly stronger than the parity $(-1)^{N_F}$: the vertex can absorb an excess of $\tpsi$ but never an excess of $\psi$. For the fermion bilinears this gives
\begin{align} \label{Bilinears}
    \langle\psi^i\psi^j\rangle_{\text{SUSY}} = 0\,, \qquad \langle\tpsi^i\psi^j\rangle_{\text{SUSY}} = R^{ij}\,, \qquad \langle\tpsi^i\tpsi^j\rangle_{\text{SUSY}} \propto \fs\,.
\end{align}
The asymmetry between $\langle\psi\psi\rangle = 0$ and $\langle\tpsi\tpsi\rangle\neq0$ is where the violation of the fluctuation-dissipation theorem is encoded. 
Any observable built from the physical fields $\phi$ and $\hat\phi$ alone has the same average in the two theories, since the vertex only raises ghost number, provided the MSR and $N\!=\!1$ path integrals are evaluated with the same regulator and ghost-number-neutral endpoint prescription. This includes the stationary or infinite-time causal prescription used below; finite-time fermionic boundary states can carry ghost number and then require separate treatment.

\subsubsection*{\ul{\it Causality and probability conservation}}

Causality is recovered from $Q_\chi$ and the ghost selection rule discussed above. Acting with \C{SUSY-delta-T} on $\mathcal O=\phi^i(t)\psi^j(t')$, the Ward identity \C{Ward-NoObstruction} reads
\begin{align} \label{N1-causality-ward}
    0
    = \big\langle\delta_\chi[\phi^i(t)\psi^j(t')]\big\rangle
    = -\epsilon\,\langle\chi^i(t)\psi^j(t')\rangle
      + \frac{\epsilon}{2T}\big\langle\phi^i(t)(\dot\phi^j-f^j)(t')\big\rangle\,.
\end{align}
The second term is the response \C{Resp-Noise}, so the mixed ghost bilinear is exactly the response,
\begin{align} \label{N1-resp-ghost}
    \langle\chi^i(t)\psi^j(t')\rangle = R^{ij}(t,t')\,.
\end{align}
In the $(\psi,\tpsi)$ frame $\langle\chi^i\psi^j\rangle=\langle\tpsi^i\psi^j\rangle-T\langle\psi^i\psi^j\rangle=\langle\tpsi^i\psi^j\rangle$, since $\langle\psi\psi\rangle_{\text{SUSY}}=0$ by \C{Bilinears}. The ghost propagator is retarded, so the response cannot precede the perturbation,
\begin{align} \label{N1-causality}
    R^{ij}(t,t') = 0 \qquad (t<t')\,.
\end{align}
In the Parisi-Sourlas formulation this followed from the separately conserved BRST charge $Q$ \C{Causality}; here it follows from the single supercharge $Q_\chi$.

Probability conservation follows through the same Ward identity together with the one-sided selection rule. The response field is $Q_\chi$-exact, $\delta_\chi\psi^i=-i\epsilon\,\hat\phi^i$, but it is not $Q_\chi$-closed: $\delta_\chi\hat\phi^i=i\epsilon\,\dot\psi^i$. Applying \C{Ward-NoObstruction} to $\mathcal O=\psi^{i_1}(t_1)\hat\phi^{i_2}(t_2)\cdots\hat\phi^{i_n}(t_n)$ gives
\begin{align} \label{N1-aux-relation}
    \big\langle\hat\phi^{i_1}(t_1)\cdots\hat\phi^{i_n}(t_n)\big\rangle
    = -\sum_{a=2}^{n}
    \big\langle\psi^{i_1}(t_1)\hat\phi^{i_2}(t_2)\cdots\dot\psi^{i_a}(t_a)\cdots\hat\phi^{i_n}(t_n)\big\rangle\,.
\end{align}
Every term on the right carries two $\psi$'s and no $\tpsi$'s, so $\#\tpsi-\#\psi=-2$ and it is killed by the one-sided selection rule \C{Selection-N1}. Therefore
\begin{align} \label{N1-aux-noAA}
    \big\langle\hat\phi^{i_1}(t_1)\cdots\hat\phi^{i_n}(t_n)\big\rangle = 0\,, \qquad n\geq1\,,
\end{align}
and expanding $Z[h]=\big\langle e^{-i\int dt\,h_i\hat\phi^i}\big\rangle_{\text{SUSY}}$ gives $Z[h]=Z$ as in \C{Z-conserved}. The total probability is unchanged by an applied force. The conclusion coincides with the Parisi-Sourlas result, but the mechanism now uses $Q_\chi$ together with the sharper one-sided selection rule rather than a separately conserved BRST charge.

\subsubsection*{\ul{\it The fluctuation-dissipation relation and its violation}}

The fluctuation-dissipation relation is the identity obtained from $\mathcal O=\phi^i(t)\chi^j(t')$. Using $\delta_\chi\phi^i=-\epsilon\chi^i$ and $\delta_\chi\chi^i=-\epsilon\dot\phi^i$, \C{Ward-NoObstruction} gives
\begin{align} \label{N1-fdt-ward}
    0 = \big\langle\delta_\chi[\phi^i(t)\chi^j(t')]\big\rangle
    = -\epsilon\Big(
        \langle\chi^i(t)\chi^j(t')\rangle
        + \langle\phi^i(t)\dot\phi^j(t')\rangle
    \Big)\,.
\end{align}
Thus the single supercharge implies the source-free identity
\begin{align} \label{FDT-chi}
    \langle\chi^i(t)\,\chi^j(t')\rangle = -\,\p_{t'}C^{ij}(t,t')\,,
\end{align}
valid for any force. Decomposing $\chi^i=\tpsi^i-T\psi^i$ \C{T-result} and using \C{Bilinears}, $\langle\chi^i\chi^j\rangle=\langle\tpsi^i\tpsi^j\rangle_{\text{SUSY}}-T\big(R^{ij}(t,t')-R^{ji}(t',t)\big)$, which turns \C{FDT-chi} into
\begin{align} \label{FDT-Violation}
    R^{ij}(t,t') - R^{ji}(t',t) - \frac{1}{T}\,\p_{t'}C^{ij}(t,t') = \Sigma^{ij}\,, \qquad \Sigma^{ij} = \frac{1}{T}\,\langle\tpsi^i(t)\,\tpsi^j(t')\rangle_{\text{SUSY}}\,.
\end{align}
This is the exact analog of the Parisi-Sourlas relation \C{FDT-alltime}, but now the violation is not an obstruction to a broken symmetry. The single unbroken supercharge $Q_\chi$ supplies the exact identity \C{FDT-chi}, and the missing piece of the equilibrium relation is promoted to the exact value of the response-ghost condensate $\langle\tpsi^i\tpsi^j\rangle_{\text{SUSY}} = T\,\Sigma^{ij}$. In this sense the fluctuation-dissipation theorem is not so much violated as superseded by a universal relation, valid for any force, that reduces to fluctuation dissipation when the right-hand side of \C{FDT-Violation} vanishes.

\subsubsection*{\ul{\it Evaluating the response-ghost condensate}}

Because the ghost sector of the Parisi-Sourlas theory is quadratic, the condensate can be evaluated in closed form. The bilinear $\langle\tpsi^i\tpsi^j\rangle_{\text{SUSY}}$ has $\#\tpsi-\#\psi=2$, so the selection rule \C{Selection-N1} picks out the single-vertex term $m=1$ in \C{SUSY-Average-Expansion}:
\begin{align} \label{ttbar-vertex}
    \langle\tpsi^i(t)\tpsi^j(t')\rangle_{\text{SUSY}}
    =
    \frac{T}{2}\int dt''\,
    \big\langle
    \fs_{kl}(\phi(t''))\,
    \tpsi^i(t)\tpsi^j(t')\psi^k(t'')\psi^l(t'')
    \big\rangle_{\text{MSR}}\,.
\end{align}
The tensor $\fs_{kl}(\phi(t''))$ is a function of the fluctuating trajectory and remains inside the average. Splitting the action into bosonic and ghost parts, $S_{\text{MSR}}=S^B[\phi]+S^F[\psi,\tpsi;\phi]$ with $S^F=\int dt\,\tpsi^j(\partial_t\delta_j^i+\p_j f^i)\psi_i$, define
\begin{align} \label{ghost-average}
    \langle\cdots\rangle_{\psi\tpsi}
    = \int D\psi\,D\tpsi\,(\cdots)e^{-S^F[\psi,\tpsi;\phi]}\,,\qquad
    \langle\cdots\rangle_{\phi}
    = \int D\phi\,(\cdots)e^{-S^B[\phi]}\,.
\end{align}
The MSR average is then a quadratic ghost integral at fixed trajectory followed by the bosonic trajectory average. The only nonzero ghost contraction is the retarded propagator
\begin{align} \label{Ghost-Green}
    &G^{ik}[\phi](t,t') = \langle\tpsi^i(t)\,\psi^k(t')\rangle_{\psi\tpsi}
    = \theta(t-t')\Big[\mathcal T\exp\!\int_{t'}^t\!\p f(\phi(s))\,ds\Big]^{ik}\,, \\
    &\langle\psi\psi\rangle_{\psi\tpsi}=0\,,\qquad
    \langle\tpsi\tpsi\rangle_{\psi\tpsi}=0\,.
\end{align}
This propagator solves $\big(\delta^i_j\p_t-\p_j f^i(\phi(t))\big)G^{jk}[\phi](t,t')=\delta^{ik}\delta(t-t')$, and its average over trajectories returns the physical response, $R^{ik}=\langle G^{ik}[\phi]\rangle_\phi$. Wick contracting the four ghosts in \C{ttbar-vertex}, the two pairings combine through the antisymmetry of $\fs_{kl}$:
\begin{align} \label{ghost-wick}
    \fs_{kl}(t'')\,
    \langle\tpsi^i(t)\tpsi^j(t')\psi^k(t'')\psi^l(t'')\rangle_{\psi\tpsi}
    =
    -2\,\fs_{kl}(t'')\,G^{ik}[\phi](t,t'')\,G^{jl}[\phi](t',t'')\,.
\end{align}
Only the bosonic average remains,
\begin{align} \label{ttbar-GG}
    \langle\tpsi^i(t)\,\tpsi^j(t')\rangle_{\text{SUSY}} = -T\int dt''\,\big\langle\,\fs_{kl}(\phi(t''))\,G^{ik}[\phi](t,t'')\,G^{jl}[\phi](t',t'')\,\big\rangle_\phi\,.
\end{align}
With the state-transition matrix of the tangent flow\footnote{After linearizing the deterministic flow $\dot\phi^i=f^i(\phi)$ around a trajectory $\phi(t)$, a perturbation $v^i$ evolves as $\dot v^i=\p_j f^i(\phi(t))v^j$. Thus $v(t)=U[\phi](t,t')v(t')$, with $U(t,t)=1$ and $U(t,t')U(t',t'')=U(t,t'')$.} $U[\phi](t,t'')=\mathcal T\exp\int_{t''}^{t}\p f(\phi(s))\,ds$, the two Heaviside factors restrict the internal time to the common past and
\begin{align} \label{ttbar-explicit}
    \big\langle\tpsi(t)\,\tpsi(t')^{\text T}\big\rangle_{\text{SUSY}} = -T\int\limits_{-\infty}^{\min(t,t')} dt''\,\big\langle\,U(t,t'')\,\fs(\phi(t''))\,U(t',t'')^{\text T}\,\big\rangle_\phi\,.
\end{align}
This is the physical content of the violation: the non-reciprocity $\fs$ is switched on at an internal time $t''$ in the common past of both external times, each index is dragged forward by a retarded tangent-flow propagator, and the result is averaged over trajectories. The right-hand side of (\ref{ttbar-explicit}) gives a compact geometric evaluation of the response-ghost condensate, and hence of the fluctuation-dissipation violation; to the best of our knowledge, this representation is novel.
Such violations have been studied using different approaches in earlier work \cite{cugliandolo_energy_1997,harada_equality_2005,speck_restoring_2006,seifert_fluctuation_2010,marconi_fluctuation_2008,baiesi_fluctuations_2009,feng_potential_2011}.

\subsubsection*{\ul{\it Higher-order fluctuation-dissipation relations}}

The construction extends to every order. Decorating the $n$-point function with a single $\chi$ and using $\delta_\chi\phi^i=-\epsilon\chi^i$ and $\delta_\chi\chi^j=-\epsilon\dot\phi^j$, the source-free identity \C{Ward-NoObstruction} applied to $\mathcal O=\phi^{i_1}(t_1)\cdots\phi^{i_n}(t_n)\chi^j(t')$ gives
\begin{align} \label{FDT-npt-ward}
    \sum_{a=1}^{n}
    \big\langle
    \phi^{i_1}(t_1)\cdots\chi^{i_a}(t_a)\cdots\phi^{i_n}(t_n)\chi^j(t')
    \big\rangle
    +\p_{t'}C_{n+1}^{i_1\cdots i_n j}=0\,,
\end{align}
where $C_{n+1}^{i_1\cdots i_n j}=\langle\phi^{i_1}(t_1)\cdots\phi^{i_n}(t_n)\phi^j(t')\rangle$. Rewriting each $\chi$ as $\chi=\tpsi-T\psi$, the selection rule \C{Selection-N1} splits the left-hand side into ghost-neutral response terms and terms with two $\tpsi$'s that encode the fluctuation-dissipation violation.

For the three-point case, take $\mathcal O=\phi^i(t_1)\phi^j(t_2)\chi^k(t_3)$. Then \C{FDT-npt-ward} reads
\begin{align} \label{FDT-3pt-ward}
    \big\langle\chi^i(t_1)\phi^j(t_2)\chi^k(t_3)\big\rangle
    +\big\langle\phi^i(t_1)\chi^j(t_2)\chi^k(t_3)\big\rangle
    +\p_{t_3}C_3^{ijk}=0\,.
\end{align}
Expanding each fermion with $\chi=\tpsi-T\psi$ and discarding the two-$\psi$ terms by \C{Selection-N1}, we get
\begin{align} \label{FDT-3pt-expanded}
\begin{aligned}
    &\langle\tpsi^i(t_1)\phi^j(t_2)\psi^k(t_3)\rangle
    +\langle\psi^i(t_1)\phi^j(t_2)\tpsi^k(t_3)\rangle \\
    &+\langle\phi^i(t_1)\tpsi^j(t_2)\psi^k(t_3)\rangle
    +\langle\phi^i(t_1)\psi^j(t_2)\tpsi^k(t_3)\rangle
    -\frac{1}{T}\p_{t_3}C_3^{ijk} \\
    &=
    \frac{1}{T}\Big[
    \langle\tpsi^i(t_1)\phi^j(t_2)\tpsi^k(t_3)\rangle
    +\langle\phi^i(t_1)\tpsi^j(t_2)\tpsi^k(t_3)\rangle
    \Big]\,.
\end{aligned}
\end{align}
The ghost-neutral correlators are physical responses. At fixed trajectory,
\begin{align} \label{Ghost-Resp}
    \langle\tpsi^i(t)\psi^k(t')\rangle_{\psi\tpsi}
    = G^{ik}[\phi](t,t')
    = \frac{\delta\phi^i(t)}{\delta h_k(t')}\,.
\end{align}
Using this relation, the first and third terms on the left of \C{FDT-3pt-expanded} combine into the response of the two-point correlator to a force at $t_3$,
\begin{align}
\begin{aligned}
    &\langle\tpsi^i(t_1)\phi^j(t_2)\psi^k(t_3)\rangle
    +\langle\phi^i(t_1)\tpsi^j(t_2)\psi^k(t_3)\rangle \\
    &\qquad =
    \left\langle\frac{\delta\phi^i(t_1)}{\delta h_k(t_3)}\,\phi^j(t_2)\right\rangle
    +\left\langle\phi^i(t_1)\frac{\delta\phi^j(t_2)}{\delta h_k(t_3)}\right\rangle
    =
    \frac{\delta\langle\phi^i(t_1)\phi^j(t_2)\rangle}{\delta h_k(t_3)}\,.
\end{aligned}
\end{align}
The remaining two ghost-neutral terms are advanced responses of the marked leg, giving
\begin{align} \label{FDT-3pt-explicit}
\begin{aligned}
    &\frac{\delta\langle\phi^i(t_1)\phi^j(t_2)\rangle}{\delta h_k(t_3)}
    -\left\langle\frac{\delta\phi^k(t_3)}{\delta h_i(t_1)}\,\phi^j(t_2)\right\rangle
    -\left\langle\frac{\delta\phi^k(t_3)}{\delta h_j(t_2)}\,\phi^i(t_1)\right\rangle \\
    &\qquad\qquad
    -\frac{1}{T}\p_{t_3}\big\langle\phi^i(t_1)\phi^j(t_2)\phi^k(t_3)\big\rangle
    = \Sigma_3^{ijk}\,.
\end{aligned}
\end{align}
Here the violation is the pair of response-ghost condensates on the right of \C{FDT-3pt-expanded},
\begin{align} \label{Sigma3}
    \Sigma_3^{ijk}
    =
    \frac{1}{T}\Big[
    \big\langle\tpsi^i(t_1)\phi^j(t_2)\tpsi^k(t_3)\big\rangle_{\text{SUSY}}
    +
    \big\langle\phi^i(t_1)\tpsi^j(t_2)\tpsi^k(t_3)\big\rangle_{\text{SUSY}}
    \Big]\,.
\end{align}
Applying the same logic to $\mathcal O=\phi^{i_1}(t_1)\cdots\phi^{i_n}(t_n)\chi^j(t')$ gives
\begin{align} \label{FDT-npt}
\begin{aligned}
    &\frac{\delta\langle\phi^{i_1}(t_1)\cdots\phi^{i_n}(t_n)\rangle}{\delta h_j(t')}
    -\sum_{a=1}^{n}
    \Big\langle
    \frac{\delta\phi^j(t')}{\delta h_{i_a}(t_a)}
    \Big(\prod_{b\neq a}\phi^{i_b}(t_b)\Big)
    \Big\rangle
    -\frac{1}{T}\p_{t'}C_{n+1}^{i_1\cdots i_n j}
    =
    \Sigma_{n+1}^{i_1\cdots i_n j}\,.
\end{aligned}
\end{align}
The violation is
\begin{align} \label{Sigma-n}
    \Sigma_{n+1}^{i_1\cdots i_n j} = \frac{1}{T}\sum_{a=1}^{n}\big\langle\phi^{i_1}(t_1)\cdots\tpsi^{i_a}(t_a)\cdots\phi^{i_n}(t_n)\,\tpsi^j(t')\big\rangle_{\text{SUSY}}\,,
\end{align}
which is the higher-rank generalization of the response-ghost condensate. Each term has two $\tpsi$'s, one on the marked leg $j$ at $t'$ and one on the promoted leg $a$, so by \C{Selection-N1} it survives through a single vertex. Integrating out the Gaussian ghosts as before gives
\begin{align} \label{Sigma-n-GG}
\begin{aligned}
    &\big\langle
    \phi^{i_1}(t_1)\cdots\tpsi^{i_a}(t_a)\cdots\phi^{i_n}(t_n)\tpsi^j(t')
    \big\rangle_{\text{SUSY}} \\
    &\qquad =
    -T\int dt''\,
    \Big\langle
    \Big(\prod_{b\neq a}\phi^{i_b}(t_b)\Big)
    \fs_{kl}(\phi(t''))\,
    G^{i_a k}[\phi](t_a,t'')\,
    G^{jl}[\phi](t',t'')
    \Big\rangle_\phi\,.
\end{aligned}
\end{align}
Equivalently,
\begin{align} \nonumber
\begin{aligned}
    \Sigma_{n+1}^{i_1\cdots i_n j}
    =
    -\sum_{a=1}^{n}
    \int_{-\infty}^{\min(t_a,t')} dt''\,
    \Big\langle
    \Big(\prod_{b\neq a}\phi^{i_b}(t_b)\Big)
    U^{i_a k}[\phi](t_a,t'')\,
    \fs_{kl}(\phi(t''))\,
    U^{jl}[\phi](t',t'')
    \Big\rangle_\phi\,.
\end{aligned}
\end{align}
At each order the non-reciprocity tensor is inserted in the common past of the marked leg and one promoted leg, its two indices are propagated forward by the retarded tangent flow, and the remaining fields spectate. For $n=1$ this reduces to the two-point condensate \C{ttbar-explicit}, while $n=2$ gives the three-point relation \C{FDT-3pt-explicit}.

\section{Stochastic PDEs and Quantum Field Theory}

\label{Sec-SPDE}

\subsection{The reciprocal case and Parisi-Sourlas supersymmetry}

We can generalize our preceding discussion to higher dimensions. The natural generalization of the Langevin equation (\ref{Langevin-General}) is a stochastic partial differential equation (SPDE). For example a generalization of the reciprocal theory of (\ref{Langevin-General}) where $f^i = \partial^i V$  is given by
\begin{align} \label{SPDE-Rec}
    \begin{aligned}
    &\partial_t \phi^i(\vec x,t) = - \frac{\delta S[\phi(\vec x,t)]}{\delta \phi^i(\vec x,t)}  + \xi^i(\vec x,t)\,.
    \end{aligned}
\end{align}
Here the action $S[\phi(\vec x, t)]$ is a functional of field configurations which takes the form\footnote{This action is not to be confused with the MSR action.}
\begin{align} \label{freeenergy}
    S[\phi(\vec x, t)] = \int_{\Sigma^d} d^dx\, \mathcal{L}(\phi(\vec x, t))\,.
\end{align}
Note that time is just a label in \C{freeenergy} with integration only over the spatial manifold $\Sigma^d$. There are no time derivatives of fields appearing on the right hand side of \C{freeenergy}. 
Fields are maps,
\begin{align}
    \phi^i(\vec x, t) : \Sigma^d \times \mathcal{I} \rightarrow M^n\,,
\end{align}
where $M$ is the target manifold and $t$ takes values in the interval $\mathcal{I}$. For simplicity, we again take a flat metric $g_{ij} = \delta_{ij}$ for $M$.\footnote{This special class of SPDEs can be generalized to curved target spaces using the same procedure described in (\ref{Langevin-Colored}) of Appendix~\ref{App-Ito-Stratonovich}.} Sometimes $\Sigma^{d+1}=\Sigma^d \times \mathcal{I}$ is called spacetime with the caveat that time refers to Euclidean time.

We consider Gaussian white noise, $\xi^i(\vec x, t)$, on spacetime with the partition function,
\begin{align}
    &Z^{d+1} = \int D \xi_i(\vec x,t) \exp\left(-\frac{1}{2\sigma}\int_{\mathcal{I} \times \Sigma^d} dt d^d x \ \xi_i^2(\vec x,t)\right)\,,
\end{align}
normalized such that
\begin{align}
    \langle \xi_i(\vec x,t) \xi_j(\vec x',t') \rangle = \sigma \delta_{ij}\delta(t-t') \delta^d(\vec x - \vec x')\,.
\end{align}
Here $\sigma$ is a measure of the noise strength, which will be set to one from now on. This convention fixes the normalization of the eventual equilibrium weight: for general $\sigma$, the zero-current condition in the Fokker-Planck equation gives $P_\star\propto \exp(-2S/\sigma)$, so the common $e^{-S}$ convention corresponds to choosing $\sigma=2$ or to absorbing this factor into $S$. A classic example is the choice 
\begin{align} \label{phi-fourth}
        &S[\phi(\vec x)] = \int_{\Sigma^d} d^d x \left[ \frac{\mathcal{D}}{2} (\nabla \phi)^2 + \frac{m^2}{2} \phi^2 + \frac{\lambda}{4} \phi^4 \right]\,,
\end{align}
which leads to the SPDE
\begin{align} \label{Kin-Ising}
        \left(\partial_t - \mathcal{D}\nabla^2\right) \phi + m^2 \phi + \lambda \phi^3 =   \xi\,.
\end{align}
This is stochastic quantization of $\phi^4$ field theory on $\Sigma^d$ discussed, for example, in \cite{Damgaard:1987rr,hairer_theory_2014}. 

Note that (\ref{phi-fourth}) describes the Ising model on a lattice that discretizes $\Sigma^d$ in a suitable continuum limit. Therefore (\ref{Kin-Ising}) should describe the time-dependent behaviour of this system with some choice of spin dynamics. This is indeed correct if one assigns Glauber dynamics to the spins of the Ising model \cite{glauber_timedependent_1963,henkel_non-equilibrium_2010}. Indeed there is a rigorous proof that an Ising model with Glauber dynamics on a two-dimensional torus, $\Sigma = T^2$, is given by (\ref{Kin-Ising})~\cite{mourrat_convergence_2015}. A special case of \C{phi-fourth} is the Gaussian free theory with $m^2=\lambda=0$ with SPDE, 
\begin{align} \label{Stochastic-Heat}
        \left(\partial_t - \mathcal{D} \nabla^2\right) \phi = \xi\,,
\end{align}
which is the stochastic heat equation on $\Sigma^d$. Here $\phi(t,\vec x)$ is the temperature field at point $(\vec x, t)$ and $\mathcal{D}$ is the diffusion constant.

The flow described by (\ref{SPDE-Rec}) is a formal expression that requires both ultraviolet and possibly infrared regularization in a manner familiar from the renormalization of quantum field theory. The hope of stochastic quantization is that (\ref{SPDE-Rec}) physically describes the time evolution of $\phi^i$ toward an equilibrium configuration of the $d$-dimensional Euclidean field theory defined by \C{freeenergy}, after dropping the time label. 

If one were to blithely ignore regularization issues then the Fokker-Planck equation of (\ref{Fokker-Planck}) generalizes to field theory:
\begin{align} \label{Func-FP}
    \partial_t P[\phi(\vec x),t] = \int_{\Sigma^d} d^dx \frac{\delta}{\delta \phi^i(\vec x)}\left( \frac{\delta}{\delta \phi_i(\vec x)}\frac{1}{2} P[\phi(\vec x),t] + \frac{\delta S[\phi(\vec x)]}{\delta \phi_i(\vec x)} P[\phi(\vec x),t] \right)\,.
\end{align}
The stationary distribution is given by,
\begin{align}
    P_\star[\phi(\vec x)] = \frac{1}{Z^d} \exp\left(-2 S[\phi( \vec x)]\right)\,, && Z^d = \int D\phi(\vec x) \exp\left(-2 S[\phi( \vec x)]\right)\,,
\end{align}
and the probability for a field configuration, $\phi^i(\vec x)$, in equilibrium is given by the Gibbs distribution $P_\star[\phi(\vec x)]$. However, this is all purely formal. The approach we will take is to define the stochastic system via its map to an MSR quantum field theory, discussed in Section~\ref{MSR}. This will allow us to import some intuition from formal high-energy physics. As in the quantum mechanics case, zero modes need to be treated carefully. For example, the stochastic heat equation of \C{Stochastic-Heat} has zero modes that grow linearly with a spatial coordinate but those configurations will often correspond to unphysical backgrounds; for example, if $\phi$ describes a temperature then we might restrict to configurations with a bounded $\phi$ via the choice of boundary conditions. For the quantum mechanics case, there is a proof of the perturbative equivalence of the stochastic system and the corresponding quantum system found in~\cite{bonicelli2023algebraic}.

Following this philosophy, we apply the MSR procedure to (\ref{SPDE-Rec}) giving,
\begin{align}
    S_{MSR}^{d+1} = \int\limits_{\Sigma \times \mathcal{I}} dt d^d x \left[ \frac{(\partial_t \phi_i)^2}{2} + \frac{1}{2}\left( \frac{\delta S}{\delta \phi_i} \right)^2 + \tpsi^j \left(\partial_t \delta^i_{j} - \frac{\delta^2 S}{\delta \phi^i \delta \phi_j} \right) \psi_i \right]\,.
\end{align}
This action has Parisi-Sourlas supersymmetry which is implemented by two transformations,
\begin{align} \label{PS-SUSY-QFT}
\begin{aligned}
    & \delta \phi^i = - \epsilon \tpsi^i\,, && \delta \psi_i = \epsilon\left(\partial_t \phi_i + \frac{\delta S}{\delta \phi^i} \right)\,, && \delta \tpsi^i = 0\,, \\
    & \tdelta \phi^i = \tepsilon \psi_i\,, && \tdelta \psi_a = 0\,, && \tdelta \tpsi^i = \tepsilon \left(-\partial_t \phi^i + \frac{\delta S}{\delta \phi_i} \right)\,.
\end{aligned}
\end{align}
This action can be reformulated in superspace using the superfields,
\begin{align}
    &\Phi^i = \phi^i + \psi^i \tilde \theta + \theta \tilde \psi^i + \theta \tilde \theta F^i\,,
\end{align}
and the supercovariant derivatives and supercharges,
\begin{align}
\begin{aligned}
    & D = \frac{\partial}{\partial \theta} + 2 \tilde \theta \frac{\partial}{\partial t}\,, && \qquad \tilde D = \frac{\partial}{\partial \tilde \theta}\,, \\
    & Q = - \frac{\partial}{\partial \theta}\,, && \qquad \tilde Q = -\frac{\partial}{\partial \tilde \theta} + 2 \theta \frac{\partial}{\partial t}\,,
\end{aligned}
\end{align}
as follows:
\begin{align}
    S_{MSR}^{d+1} = \int\limits\limits_{\Sigma \times \mathcal{I}} dt d^d x d \tilde\theta d \theta \left[ \frac{1}{2}D\Phi^i \tilde D \Phi_i - S[\Phi] \right]\,.
\end{align}

\subsubsection*{\ul{\it Two examples of reciprocal SPDEs and their $N=2$ SUSY formulations}}

To illustrate the correspondence, let us describe the field theories dual to the two examples \C{Kin-Ising} and (\ref{Stochastic-Heat}). Applied to the stochastic heat equation (\ref{Stochastic-Heat}) we get
\begin{align} \label{N=2-SUSY-Lifshitz}
    S_{MSR} = \int_{\Sigma^d \times \mathbb{R}} dt d^d x \left[ \frac{\left(\partial_t \phi\right)^2}{2} +\mathcal{D}^2 \frac{(\nabla^2 \phi)^2}{2} + \tpsi\left(\partial_t - \mathcal{D} \nabla^2\right) \psi\right]\,.
\end{align}
The bosonic terms are the $z=2$ Lifshitz theory, while the fermionic couplings make the theory invariant under the $N=2$ SUSY of (\ref{PS-SUSY-QFT}). It is interesting to see that the stochastic heat equation on $\Sigma^d$ is formally equivalent to the supersymmetric Lifshitz theory on $\Sigma^d \times \mathbb{R}$; see, for example,~\cite{chapman2015supersymmetric}.

For (\ref{Kin-Ising}), which describes the continuum limit of the kinetic Ising model on $\Sigma^d$, we find the quantum action
\begin{align} \label{Rec-Kin-Ising-Ac}
    S_{MSR} =  \int\limits_{\Sigma^d \times \mathbb{R}} dt d^dx \Big[ & \frac{\left(\partial_t \phi\right)^2}{2} +\frac{1}{2} \left(\mathcal{D}\nabla^2 \phi -  m^2 \phi - \lambda \phi^3\right)^2  \cr &  + \tpsi\left(\partial_t - \mathcal{D}\nabla^2 + m^2   + 3\lambda\phi^2\right)\psi \Big],
\end{align}
which is an interacting generalization of the $N=2$ Lifshitz theory.

\subsection{Nonreciprocal SPDEs and their field theory formulations}\label{Sec-NR-SPDE}

Now we turn to how this story changes if we have nonreciprocal interactions. Generalizing the discussion of Section \ref{Sec-magic-of-reciprocity}, we expect nonreciprocity to be encoded in vector fields $A_i[\phi(\vec x,t)]$ that are not gradients of some scalar functional $S[\phi(\vec x)]$. As a simple example, pick a two-field generalization of (\ref{phi-fourth}) with $O(2)$ symmetry and then add nonreciprocal couplings controlled by $(\mathcal{A}, a)$, which break the symmetry to $SO(2)$:
\begin{align}
    \left\{
    \begin{aligned}
        &(\partial_t - \mathcal{D} \nabla^2) \phi_1 + m^2 \phi_1 + \lambda \phi_1(\phi_1^2 + \phi_2^2) - \mathcal{A} \nabla^2 \phi_2 + a \phi_2 =  \xi_1\,, \\
        &(\partial_t - \mathcal{D} \nabla^2) \phi_2 + m^2 \phi_2 + \lambda \phi_2 (\phi_1^2 + \phi_2^2) + \mathcal{A} \nabla^2 \phi_1 - a \phi_1 =   \xi_2\,.
    \end{aligned}\right.
\end{align}
These equations can be written as
\begin{align}
    \left\{
    \begin{aligned}
        &\partial_t \phi_1 = - \frac{\delta S}{\delta \phi_1} - \frac{\delta \widehat{S}}{\delta \phi_2} + \xi_1\,, \\
        &\partial_t \phi_2 = - \frac{\delta S}{\delta \phi_2} + \frac{\delta \widehat{S}}{\delta \phi_1} + \xi_2\,,
    \end{aligned}
    \right.
\end{align}
where
\begin{align}
\begin{aligned}
    &S = \int d^d x \left[\frac{\mathcal{D}}{2}\big((\nabla \phi_1)^2+(\nabla \phi_2)^2\big) + \frac{m^2}{2}(\phi_1^2+\phi_2^2) + \frac{\lambda}{4} (\phi_1^2+\phi_2^2)^2 \right]\,, \\
    &\widehat{S} = \int d^d x \left[\frac{\mathcal{A}}{2}\big((\nabla \phi_1)^2+(\nabla \phi_2)^2\big) + \frac{a}{2}(\phi_1^2+\phi_2^2) \right]\,.
\end{aligned}
\end{align}
This gives an explicit field-theory counterpart of the nonreciprocal potential $U$ in \C{defA}. Choosing the orientation convention $\star d\phi_1=d\phi_2$, $\star d\phi_2=-d\phi_1$ in the two-dimensional target space with coordinates $(\phi_1, \phi_2)$ we can define the two-form
\begin{align}
    U = -\widehat{S}\,d\phi_1\wedge d\phi_2\,,
\end{align}
whose co-exact one-form is
\begin{align}
    \delta U = \frac{\delta \widehat{S}}{\delta \phi_2}\,d\phi_1-\frac{\delta \widehat{S}}{\delta \phi_1}\,d\phi_2\,.
\end{align}
Thus the non-gradient vector field in this example is
\begin{align}
    A^1=\frac{\delta \widehat{S}}{\delta \phi_2}\,, \qquad A^2=-\frac{\delta \widehat{S}}{\delta \phi_1}\,,
\end{align}
or, equivalently, $A^i=\varepsilon^{ij}\delta\widehat{S}/\delta\phi_j$ with $\varepsilon^{12}=1$. The scalar functional $\widehat{S}$ is therefore not added to $S$ as another reciprocal free energy; its functional derivative is first rotated by the antisymmetric tensor in field-component space, giving a co-exact rather than exact contribution to the force.

In accord with the intuition of Section \ref{Sec-magic-of-reciprocity}, the nonreciprocity is encoded in the non-gradient component of the force. For the general nonreciprocal SPDE, we write
\begin{align} \label{SPDE-NR}
    \partial_t \phi^i = - \delta^i S(\phi) - A^i(\phi) + \x^i\,,
\end{align}
where $\delta^i$ is a shorthand for $\frac{\delta}{\delta\phi^i}$. Again it is sufficient that 
\begin{align}
    \delta_i A_j(\phi) - \delta_j A_i(\phi) \neq 0\,,
\end{align}
to have nonreciprocity. 
Just like the case of (\ref{Breaking-PS}), the MSR description of (\ref{SPDE-NR}) does not have Parisi-Sourlas supersymmetry. However, we can again use (\ref{New-Regularization}) to obtain a supersymmetric field theory for (\ref{SPDE-NR}). Using the superfields,
\begin{align}
    &\Lambda^i = \psi^i + \theta F^i\,, && \varphi^i = \phi^i + \theta \chi^i\,,
\end{align}
and SUSY covariant derivative and supercharge
\begin{align}
    &D_\chi = - \partial_\theta - \theta \partial_t\,, && Q_\chi = - \partial_\theta + \theta \partial_t\,.
\end{align}
As before $Q_\chi^2=-\partial_t$, $D_\chi^2=\partial_t$ and $\{Q_\chi,D_\chi\} = 0$. The supersymmetric quantum field theory dual to (\ref{SPDE-NR}) is given by
\begin{align} \label{FT-Action-Superspace}
    S_{SUSY} & = \int dt d^dx d\theta \left(\frac{1}{2} \Lambda^i D_\chi \Lambda_i - \Lambda^i E_i(\varphi) \right)\,,
\end{align}
with
\begin{align}
    E^i(\phi) = \partial_t \phi^i + \delta^i S(\phi) + A^i(\phi)\,.
\end{align}
Integrating out the auxiliary variable $F^i$ gives,
\begin{align} \label{FT-Action}
    S_{SUSY} =&  \int dtd^dx \bigg[ \frac{\left(\partial_t \phi^i\right)^2}{2} + \partial_t \phi^i A_i + \frac{\left(\delta^i S + A^i\right)^2}{2}
    \cr & \qquad \qquad +\frac{1}{2} \psi^i \partial_t \psi_i + \chi^i \partial_t \psi_i - \chi_j \left(\delta_j \delta^i S + \delta_j A^i\right) \psi_i \bigg]\,,
\end{align}
which is invariant under the symmetry:
\begin{align} \label{Q-chi-coord-QFT}
    & \delta_\chi \phi^i = - \epsilon \chi^i\,, && \delta_\chi \psi_i = \epsilon \left(\partial_t \phi_i + \delta_i S + A_i\right)\,, && \delta_\chi\chi^i = - \epsilon \partial_t \phi^i\,.
\end{align}
The supercharge from the Noether procedure is,
\begin{align}
    &Q_\chi = \chi^i ( \partial_t \phi_i + \delta_i S + A_i ) + \psi_i \left( \delta^i S + A^i\right)\,.
\end{align}
The canonical momenta take the form,
\begin{align}
\begin{aligned}
    &p_i = i\frac{\partial \mathcal{L}}{\partial( \partial_t \phi^i)} = i\left(\partial_t \phi_i + A_i\right)\,,  \qquad p_i^\psi = i\frac{\partial \mathcal{L}}{\partial (\partial_t \psi^i)} = - \frac{i}{2}(\psi_i + \chi_i )\,,  \\ &  p_i^\chi = i\frac{\partial \mathcal{L}}{\partial (\partial_t \chi^i)} = - \frac{i}{2} \psi_i\,,
\end{aligned}
\end{align}
with the same magnetic field-like coupling appearing in $p_i$.
Following the same steps as before we find the Hamiltonian density of our system,
\begin{align}
     \mathcal{H} = \frac{(p_i -i A_i)^2}{2} + \frac{(\delta_i S + A_i )^2}{2} - \frac{\delta_j \delta_i S + \delta_j A_i}{2} [ \chi^j , \psi^i ]\,,
\end{align}
with a supercharge expressed in terms of momenta,
\begin{align}
    Q_\chi
    & = \chi^i ( -i \pi_i + \delta_i S + A_i)  + \psi^i ( \delta_i S + A_i)\,.
\end{align}
The supersymmetry algebra is $Q_\chi^2 = \mathcal{H}$. 

We can again redefine our fermionic fields to $(\tpsi^i,\psi^i)$ where $\tpsi^i = \chi^i + \frac{\psi^i}{2}$ and which results in the Hamiltonian density,
\begin{align}
    \mathcal{H} = \frac{(p_i -i A_i)^2}{2} + \frac{(\delta_i S + A_i )^2}{2} - \frac{\delta_j \delta_i S + \delta_j A_i}{2} [ \tpsi^j , \psi^i ] + \frac{\delta_j A_i-\delta_i A_j}{4} \psi^j \psi^i\,,
\end{align}
and supercharge
\begin{align}
    Q_\chi = Q + \frac{\tilde Q}{2}\,,
\end{align}
where $Q$ and $\tilde Q$ are given by,
\begin{align}
    Q= \tpsi^i \left( -i\pi_i + \delta_i S + A_i \right)\,, \qquad \tilde Q = \psi^i\left( i\pi_i  + \delta_i S + A_i \right)\,,
\end{align}
and $\pi_i = i\partial_t \phi_i$. This relates the supercharge $Q_\chi$ to the MSR nilpotent BRST charge $Q$. Finally we can write the supersymmetric action in terms of these fermions,
\begin{align} \label{SUSY-Act-NR}
    S_{SUSY} = \int dtd^dx &\left[ \frac{\left(\partial_t \phi^i\right)^2}{2} + \partial_t \phi^i A_i + \frac{\left(\delta_i S + A_i \right)^2}{2}  \right. \cr & \left.\quad +\tpsi^i \partial_t \psi_i - \tpsi^j \left(\delta_j \delta^i S + \delta_j A^i\right) \psi_i + \frac{1}{2} \psi^j \delta_j A_i \psi^i  \right]\,.
\end{align}
The last term of \C{SUSY-Act-NR} differentiates $S_{SUSY}$ from the MSR action. The two actions coincide if $A_i=0$ as expected. 

\subsubsection*{\ul{\it Two nonreciprocal stochastic PDEs and their SUSY counterparts}}

A particularly simple nonreciprocal stochastic PDE is given by,
\begin{align}
\begin{aligned}
    &\partial_t \phi_1 = \mathcal{D}\nabla^2 \phi_1 + \mathcal{K} \nabla^2 \phi_2 + \xi_1\,,\\
    &\partial_t \phi_2 = \mathcal{D}\nabla^2 \phi_2 - \mathcal{K} \nabla^2 \phi_1 + \xi_2\,.
\end{aligned}
\end{align}
The terms proportional to $\mathcal{D}$ lead to diffusion while the terms proportional to $\mathcal{K}$ lead to self-sustaining waves with dynamical critical exponent $z=2$ \cite{van2024soft}. Let us focus on the $\mathcal{D}=0$ case where self-sustaining waves appear without diffusion. The field theory description of this SPDE has action, 
\begin{align}
    S_{SUSY} = & \int\limits_{\Sigma^d \times \mathbb{R}}  dt d^dx \bigg[\frac{1}{2}\left(\dot\phi_1^2 + \dot\phi_2^2 \right) - \mathcal{K}\left(\dot\phi_1 \nabla^2 \phi_2 - \dot\phi_2 \nabla^2 \phi_1\right) + \frac{\mathcal{K}^2}{2} \left(\left(\nabla^2\phi_1\right)^2+\left(\nabla^2\phi_2\right)^2\right) \nonumber \\
    &+ \tpsi_1\dot \psi_1 + \tpsi_2\dot \psi_2 - \mathcal{K}\left(\tpsi_1 \nabla^2 \psi_2 - \tpsi_2 \nabla^2 \psi_1\right) - \frac{\mathcal{K}}{2}\big(\psi_2 \nabla^2 \psi_1 - \psi_1 \nabla^2 \psi_2 \big) \bigg]\,,
\end{align}
which is symmetric under
\begin{align} \label{Q-chi-coord-waves}
\begin{aligned}
    & \delta_\chi \phi_1 = - \epsilon \chi_1\,, && \delta_\chi \psi_1 = \epsilon \left(\partial_t \phi_1 - \mathcal{K}\nabla^2 \phi_2 \right)\,, && \delta_\chi\chi_1 = - \epsilon \partial_t \phi_1\,, \\
    & \delta_\chi \phi_2 = - \epsilon \chi_2\,, && \delta_\chi \psi_2 = \epsilon \left(\partial_t \phi_2 + \mathcal{K}\nabla^2 \phi_1\right)\,, && \delta_\chi\chi_2 = - \epsilon \partial_t \phi_2\,.
\end{aligned}
\end{align}
The $-\mathcal{K}\left(\dot\phi_1 \nabla^2 \phi_2 - \dot\phi_2 \nabla^2 \phi_1\right)$ term can be viewed as a kind of magnetic coupling between two supersymmetric Lifshitz theories of (\ref{N=2-SUSY-Lifshitz}) type. This coupling explicitly breaks $\tdelta$ in (\ref{PS-SUSY-QFT}) destroying the Parisi-Sourlas SUSY; however, by suitably modifying the fermionic terms we can still generate a theory which enjoys $N=1$ supersymmetry.

Another example is coupling the continuum description of two kinetic Ising models together with a nonreciprocal interaction via\footnote{There are many possible ways to couple two kinetic Ising models nonreciprocally. See \cite{Weiderpass2025PRE,avni2025nonreciprocalPRL} for a discussion of this point.}
\begin{align}
\begin{aligned}
    &(\partial_t - \nabla^2)\phi_1 - \mathcal{K} \nabla^2 \phi_2 + m^2 \phi_1 + \lambda \phi_1^3 = \xi_1\,,\\
    &(\partial_t - \nabla^2) \phi_2 + \mathcal{K} \nabla^2 \phi_1 + m^2 \phi_2 + \lambda \phi^3_2 = \xi_2\,.
\end{aligned}
\end{align}
The $N=1$ supersymmetric action for this model is
\begin{align}
    S_{SUSY}  = &\int\limits_{ \Sigma^d \times \mathbb{R}} dt d^d x \bigg[ \frac{\left(\partial_t \phi_1\right)^2}{2}+\frac{\left(\partial_t \phi_2\right)^2}{2} -\mathcal{K}\dot\phi_1\nabla^2\phi_2+\mathcal{K}\dot\phi_2\nabla^2\phi_1 \nonumber \\
    &+\frac{1}{2} \left(\nabla^2 \phi_1 + \mathcal{K}\nabla^2 \phi_2 - m^2 \phi_1 - \lambda \phi_1^3\right)^2+\frac{1}{2} \left(\nabla^2 \phi_2 - \mathcal{K} \nabla^2 \phi_1 - m^2 \phi_2 - \lambda \phi_2^3\right)^2 \nonumber \\
    &+  \tpsi_1\left(\partial_t - \nabla^2 + m^2+3\lambda\phi_1^2\right)\psi_1+  \tpsi_2\left(\partial_t - \nabla^2 + m^2+3\lambda\phi_2^2\right)\psi_2 \nonumber \\
    &-\mathcal{K}\left(\tpsi_1 \nabla^2 \psi_2 - \tpsi_2 \nabla^2 \psi_1\right) - \frac{\mathcal{K}}{2}\big(\psi_2 \nabla^2 \psi_1 - \psi_1 \nabla^2 \psi_2 \big) \bigg]\,.
\end{align}
This action is two copies of (\ref{Rec-Kin-Ising-Ac}) with a nonreciprocal extension, but with fermionic interactions that preserve $N=1$ supersymmetry. One can build many nonreciprocal models which are variations of this type of construction.

\section{Discussion and outlook} \label{Sec-Discussion}

The main result of this paper is that the loss of reciprocity does not remove all supersymmetry from the stochastic theory. For Langevin systems with additive Gaussian white noise and flat target-space metric, the standard MSR determinant admits an alternative real-fermion Pfaffian representation. In that representation the path integral can be written in a superspace with one real Grassmann coordinate, and the resulting action has a manifest supercharge $Q_\chi$ satisfying $Q_\chi^2=-\partial_t$ in the Lagrangian formulation, or $Q_\chi^2=H$ in the Hamiltonian formulation. Reciprocity is therefore not the condition for the existence of any supersymmetry. Rather, reciprocity is the condition for the enhancement to the Parisi-Sourlas structure with two real supercharges.

The same statement applies to the SPDEs considered in Section \ref{Sec-NR-SPDE}, with the usual field-theoretic caveat that the continuum theories must be regulated. The input is a specified stochastic evolution equation, together with a choice of noise normalization, regulator and boundary prescription. Given such an input, the MSR construction produces a path integral, and the Pfaffian rewriting used in \C{FT-Action-Superspace} gives a $Q_\chi$-supersymmetric formulation of that same regularized theory.

The unbroken $Q_\chi$ symmetry has observable consequences. The Ward identities derived in Sections \ref{Sec-PS-ward} and \ref{Sec-N1-ward} show that the ordinary fluctuation-dissipation theorem is recovered when the nonreciprocal curvature $\mathcal F_{ij}$ vanishes, while for $\mathcal F_{ij}\neq0$ the failure of the reciprocal fluctuation-dissipation relation is measured by the response-ghost condensate $\langle\tpsi^i(t)\tpsi^j(t')\rangle_{\text{SUSY}}$. This gives a precise sense in which the $N=1$ theory completes, rather than simply discards, the Parisi-Sourlas Ward identity: the fluctuation dissipation violation term becomes a computable correlator in the supersymmetric theory.

Finally, the superspace formulation constrains quantum corrections. If perturbation theory is defined with a regulator that preserves the superspace form of \C{FT-Action-Superspace}, counterterms can renormalize the functions and operators appearing in the superspace action, but they cannot explicitly break $Q_\chi$. Spontaneous supersymmetry breaking is still possible. Understanding when this happens, and whether it is tied to non-equilibrium phenomena such as exceptional points, possibly at criticality, is a natural direction for future work.

\section*{Acknowledgements}

We would like to thank Peter Littlewood, Igor Ovchinnikov, Vincenzo Vitelli and Cheyne Weis for helpful discussions and comments. G.~W. and S.~S. are supported in part by NSF Grant No. PHY2014195 and NSF Grant No. PHY2412985. G.~W. is supported in part by the Sidney Bloomenthal Fellowship at the University of Chicago.

\appendix\newpage

\section{Regularizing Stochastic Differential Equations (SDEs)}\label{App-Ito-Stratonovich}

In this section we briefly review some basics of stochastic calculus. For good references on this topic, see \cite{gardiner2004handbook,Ovchinnikov:2015rkh,kunita1990stochastic,hsu2002stochastic,Friz2020}.

\subsection{Common choices of regularization} \label{App-regularization}

Stochastic differential equations of the type (\ref{Langevin-General}) require regularization to make sense of the noise term.  A simple way to understand this necessity is to look at a slight generalization of (\ref{Langevin-General}) which includes colored noise
\begin{align} \label{Langevin-Colored}
    \dot x^i(t) = f^i\left(x(t)\right) + \sigma_a^i\left(x(t)\right)\circ_\alpha\xi^a(t)\,.
\end{align}
One can allow $f^i$ and $\sigma^i$ to depend explicitly on time and the subsequent discussion is essentially unchanged. Imagine the coordinates $x^i$ with $i=1, \ldots, n$ parametrize a Riemannian manifold $M$ with metric $g_{ij}(x)$. We will define and explain the remaining elements in this equation in a moment, but first we want to point out the deeper reason why regularization is required. This equation is really a formal expression because $\xi^a(t)$ is a distribution-valued random variable. Since it acts as a forcing term in the equation,  $x^i(t)$ also becomes a distribution. The product of distributions requires a choice of regularization and this choice is denoted by the symbol $\circ_\alpha$.

To regularize this theory we start by recasting it as an equation in terms of differentials,
\begin{align} \label{Langevin-Differential}
    dx^i(t) = f^i\left(x(t)\right) dt + \sigma_a^i\left(x(t)\right)\circ_\alpha dW^a(t)\,,
\end{align}
where $dW^a(t) = \xi^a(t) dt$ which makes $W^a(t) = \int_0^t \xi^a(s) ds$ a Wiener process; namely, a stochastic process with the properties
\begin{align} \label{Weiner}
    &\langle W^a(t) \rangle = 0\,, \qquad \left\langle W^a(t) W^b(s)\big) \right\rangle = \delta^{ab}\, \text{min}(t,s)\,.
\end{align}
These correlators follow from \C{noisecorrel} and give, for example, that
\begin{align} \label{Wiener2}
    \left\langle\big(W^a(t)-W^a(s)\big)\big(W^b(t)-W^b(s)\big)\right\rangle = \delta^{ab}|t-s|\,.
\end{align}
We can now define integration over $dW^a(t)$. We discretize time into $N$ intervals of size $\Delta t$ and label the different time instants by $t_n$, $n=1,\ldots, N$. The $\alpha$ stochastic integration over $dW^i(t)$ is then defined by,
\begin{align} \label{Integration-Theory}
    &\int_0^t G_a(x(t)) \circ_{\alpha} dW^a(t) \equiv \underset{N \to \infty}{\text{ms-lim}} \sum_{n =1}^N G_a\left( x_{\alpha,n}\right) \Delta W^a_n\,,
\end{align}
where $x_n^i = x^i(t_n)$, $\Delta W_n = W(t_{n+1}) - W(t_n)$ and the $\alpha$ midpoint is $x_{\alpha,n}^i = \alpha x_{n+1}^i + (1-\alpha) x_n^i$. The notation $\text{ms-lim}$ is the mean square limit. A sequence of random variables, $X^N$, indexed by $N$ has a mean square limit if
\begin{align}
    \lim_{N\rightarrow\infty} \int d\omega \, P(\omega) \Big[X^N(\omega) - X(\omega)\Big]^2 = \lim_{N\rightarrow\infty} \left\langle \big(X^N - X\big)^2 \right\rangle = 0\,,
\end{align}
for some $X$. If this is the case we write \begin{align}
    \underset{N \to \infty}{\text{ms-lim}}\, X^N = X\,.
\end{align}

Now we can examine the remaining quantities in \C{Langevin-Differential}. The collection of vector fields $f^i(x)$ and $\sigma_a^i(x)$ are elements of $T_xM$. We will assume that $a=1,\ldots, n$; if there are less $W^a$ than coordinates $x^i$ then some equations are purely deterministic. If there are more $W^a$ than coordinates $x^i$, the system is typically over-constrained. For the same reason we assume the rank of $\sigma$ is $n$. For uncorrelated noise satisfying \C{Weiner} we want $W^a$ to define orthonormal directions in $T_xM$.    
From the vectors $\sigma_a^i(x)$ we define the diffusion or covariance matrix\footnote{Other common names include the diffusion co-metric, noise co-metric, or the metric induced by the diffusion.}
\begin{align} \label{Diffusion-Matrix}
    \sum_{a=1}^n \sigma^i_a(x) \sigma^j_a(x) = D^{ij}(x)\,.
\end{align}
The matrix $D^{ij}(x)$ is positive definite and can be interpreted as a metric on $M$, but this is not necessarily the metric $g^{ij}(x)$. In the specific case where $\sigma^i_a(x) = e_a^i(x)$ where $e_a^i(x)$ define an orthonormal frame for $T_xM$ then 
\begin{align}
    \delta^{ab} e^i_a(x) e^j_b(x)=g^{ij}(x)\,, \qquad
    g_{ij} e^i_a(x) e^j_b(x)=\delta_{ab} \,.
\end{align}

For the family of theories (\ref{Integration-Theory}) labeled by $\alpha$, the most important choices are $\alpha=0$ which is the Ito integral and $\alpha=1/2$ which is the Stratonovich integral. It is worth mentioning that for each $\alpha$, there is a distinct definition for the derivative but different choices can be mapped to one another~\cite{gardiner2004handbook}. This can be easily seen by studying a differentiable function $F(x)$. We note that
\begin{align}
\begin{aligned}
    F(x_{n+1}) & = F(x_{\alpha,n}) + (1-\alpha) \partial_i F (x_{\alpha,n}) \Delta x^i_n + \frac{(1-\alpha)^2}{2}\partial_i \partial_j  F (x_{\alpha,n}) \Delta x^i_n \Delta x^j_n +\ldots\,,\\
    F(x_{n}) & = F(x_{\alpha,n}) -\alpha \partial_i F (x_{\alpha,n}) \Delta x^i_n + \frac{\alpha^2}{2}\partial_i \partial_j  F (x_{\alpha,n}) \Delta x^i_n \Delta x^j_n +\ldots \,,
\end{aligned}
\end{align}
where we used $x_{n+1}^i = x_{\alpha,n}^i + (1-\alpha) \Delta x_n^i$ and $x_{n}^i = x_{\alpha,n}^i -\alpha \Delta x_n^i$. Taking the difference gives,
\begin{align} \label{Delta-F_n}
    \Delta F_n = \partial_i F (x_{\alpha,n}) \Delta x^i_n + \left(\frac{1}{2} - \alpha \right) \partial_i \partial_j  F (x_{\alpha,n}) \Delta x^i_n \Delta x^j_n+\mathcal{O}(\Delta x_n^3)\,.
\end{align}
Finally since $\Delta x^i_n = f^i(x_{\alpha,n}) \Delta t_n + \sigma^i_a(x_{\alpha,n}) \Delta W^a_n$ using (\ref{Wiener2}) and (\ref{Diffusion-Matrix}), we see that
\begin{align}
    \langle \Delta x^i_n \Delta x^j_n \rangle = D^{ij}(x_{\alpha,n}) \Delta t_n + \mathcal{O}(\Delta t_n^2)\,.
\end{align}
This tells us that terms of order $(\Delta x)^2$ cannot be ignored when we are dealing with stochastic variables. Approximating $\Delta x^i_n \Delta x^j_n \approx D^{ij}(x_{\alpha,n}) \Delta t_n + \mathcal{O}(\Delta t_n^2)$ and plugging back into (\ref{Delta-F_n}) while taking the limit $\Delta t_n \rightarrow 0$, we see that the differential of $F(x)$ is given by
\begin{align} \label{Stochastic-Differential}
    d F(x)=\partial_i F d x^i+\left(\frac{1}{2}-\alpha\right) D^{ij}(x) \partial_i \partial_j F d t\,.
\end{align}
From a physicist's perspective, the most natural integration and differentiation theory for stochastic systems is therefore the Stratonovich regularization with $\alpha=1/2$. In this case, the standard rules of calculus apply which makes this choice compatible with the conventional action of diffeomorphisms of the target manifold $M$. That is, if we re-parameterize the manifold $M$ using a new coordinate system $y^i(x)$ then for Stratonovich the normal chain rule, $dy^i = \partial_j y^i d x^j$, applies and the stochastic equation for $dy^i$ is
\begin{align}
    dy^i = \partial_j y^i\left( f^j dt + \sigma_a^j\circ_{1/2} dW^a \right)\,.
\end{align}
On the other hand, for the Ito calculus we get
\begin{align}
    d y^i=\left( \partial_j y^i f^j+\frac{1}{2} D^{jk} \partial_j \partial_k y^i\right) d t+\left(\partial_j y^i \sigma^j_a\right) \circ_{0} d W^a(t)\,,
\end{align}
so diffeomorphisms do not act in the conventional manner. Ito regularization on the other hand preserves Ito isometries \cite{gaeta2004symmetrystochasticequations, Friz2020}. Ito isometries are $O(n)$ transformations acting on both $\sigma_a^i(x)$ and the noise via 
\begin{align} \label{Ito-Transformation}
    &\bar \sigma_a^i(x) = R_a^b(x) \sigma_b^i(x)\,, \qquad \qquad \bar \xi^a(t) = R_b^a(x) \xi^b(t
    )\,,
\end{align}
such that the diffusion tensor $D^{ij}(x) = \sum_{a=1}^m \sigma_a^i(x) \sigma_a^j(x)$ and the integrand of the path integral (\ref{NoiseMeasure}) are invariant
\begin{align}
    \sum_{a=1}^n\bar \sigma_a^i(x) \bar \sigma_a^j(x) = \sum_{a=1}^n \sigma_a^i(x) \sigma_a^j(x)\,, \qquad \sum_{a=1}^n\bar \xi^a(t) \bar \xi^a(t) = \sum_{a=1}^n \xi^a(t) \xi^a(t)\,.
\end{align}
We are again suppressing any explicit time-dependence in the rotation matrix $R$ but explicit time-dependence is permitted. 

It is natural to want the SDE (\ref{Langevin-Differential}) to be invariant under these transformations. To be a symmetry, we  demand that $d\bar W^a(t)$ is still a Wiener process. This requires the prescription $\Delta\bar{W}^a_n = R^a_b(x_{n})\Delta W^b_n$ so that 
\begin{align}
\begin{aligned}
    &\langle \Delta \bar W^a_n\rangle = \langle R^a_b(x_n) \Delta W^b_n\rangle = R^a_b(x_n) \langle \Delta W^b_n\rangle = 0\,, \\
    &\langle \Delta \bar W^a_n \Delta \bar W^b_n\rangle = R^a_c(x_n)R^b_d(x_n)\langle \Delta W^c_n \Delta W^d_n\rangle = R^a_c(x_n)R^b_d(x_n) \delta^{c d}\Delta t_n = \delta^{ab} \Delta t_n\,,
\end{aligned}
\end{align}
and thus $\Delta \bar W^i_n$ is a Wiener process. This singles out the Ito prescription $\alpha=0$. Only for this prescription does the noise act after the rotation. For other choices of $\alpha$, a standard calculation shows that under the Ito isometry the SDE (\ref{Langevin-Differential}) gets mapped to
\begin{align}
    d x^i = f^i dt + \bar\sigma^i_a \circ_\alpha d\bar W^a = \Big(f^i - \alpha\, \sigma^i_b R^b_a \partial_j R^a_c \sigma_c^j \Big)dt + \sigma^i_a \circ_\alpha d W^a\,.
\end{align}
This leads to a very interesting familiar situation. Before worrying about regulating (\ref{Langevin-Colored}), there were two ``classical'' symmetries
\begin{align}
    \text{Diff}(M) \times O(n)\,.
\end{align}
However when we are careful about the distributional character of this equation, we see there are two standard ways to regularize the theory: the Stratonovich regularization preserves $\text{Diff}(M)$ while $O(n)$ is violated. With Ito regularization, $\text{Diff}(M)$ is no longer a symmetry but $O(n)$ is preserved. This situation is completely analogous to the anomalies found in two-dimensional sigma models with chiral supersymmetry, like the models describing compactifications of the heterotic string.

\subsection{The Fokker-Planck equation} \label{App-FP-derivation}

To study the time evolution of a stochastic system, we can either study the time evolution of observables $F(x(t))$ which depend on $x^i(t)$ whose time evolution is given by the stochastic equation (\ref{Langevin-Colored}), or we fix the observables in time and evolve the probability distribution. For consistency, expectation values must be the same in either approach:
\begin{align} \label{COnsistency-Prob-Thy}
    \langle F(x(t)) \rangle = \int d^d x\, F(x) P(x,t|x_0,t_0)\,.
\end{align}
The Fokker-Planck equation is a time-evolution equation for the probability distribution. From the consistency condition above and the discussion in the last section it is easy to obtain the Fokker-Planck equation. 

We start with (\ref{Stochastic-Differential}) and plug in the value of $dx^i$. We get
\begin{align} \label{dF-withdxplugged}
    dF = \left( \partial_i F f^i + \left(\frac{1}{2} - \alpha\right) D^{ij} \partial_i \partial_j F\right) dt + \partial_i F \sigma^i_a \circ_{\alpha} dW\,.
\end{align}
For any $G_a$, note that the following expectation value vanishes in Ito regularization, 
\begin{align}
\begin{aligned}
    \left\langle \int_0^t G_a(x(t)) \circ_{0} dW^a(t) \right\rangle & = \underset{N \to \infty}{\text{ms-lim}} \sum_{n =1}^N \left \langle G_a\left( x_{n}\right) \big(  W^a(t_{n+1}) -  W^a(t_n) \big) \right\rangle\,, \\
    & = \underset{N \to \infty}{\text{ms-lim}} \sum_{n =1}^N \langle G_a\left( x_{n}\right)\rangle \left \langle \big(  W^a(t_{n+1}) -  W^a(t_n) \big) \right\rangle = 0\,,
\end{aligned}
\end{align}
where we used (\ref{Weiner}). This is zero because $G_a(x_n)$ does not depend on the noise at either $t_{n+1}$ or $t_n$ and can be pulled out of the expectation value. This is not true if $G_a$ is evaluated at the $\alpha \neq 0$ midpoint $x_{\alpha,n}$. To average over the noise for general $\alpha$, we first note that
\begin{align}
\begin{aligned}
    G_a\left( x_{\alpha,n}\right) \Delta W^a_n & = G_a\left( x_{n}\right) \Delta W^a_n + \alpha \partial_i G_a\left( x_{\alpha,n}\right) \Delta x^i_n \Delta W^a_n\,, \\
    & = G_a\left( x_{n}\right) \Delta W^a_n + \alpha \partial_i G_a\left( x_{n}\right) \sigma^i_a(x_{\alpha,n}) dt\,, \\
    & = G_a\left( x_{n}\right) \Delta W^a_n + \alpha \partial_i G_a\left( x_{\alpha,n}\right) \sigma^i_a(x_{\alpha,n}) dt\,.
\end{aligned}
\end{align}
That is
\begin{align} \label{Chainge-Ito-alpha}
    G_a \circ_{\alpha} dW^a = G_a \circ_{0} dW^a + \alpha \partial_i G_a \sigma^i_a dt\,.
\end{align}
For general $\alpha$ regularization, we see that
\begin{align} \label{alpha-integral-mean}
   \left\langle \int_0^t G_a(x(t)) \circ_{\alpha} dW^a(t) \right\rangle & = \alpha \int \partial_i G_a \sigma^i_a dt\,.
\end{align}
Using (\ref{Chainge-Ito-alpha}) in (\ref{dF-withdxplugged}) gives
\begin{align}
    dF = \left( \partial_i F f^i + \left(\frac{1}{2} - \alpha\right) D^{ij} \partial_i \partial_j F + \alpha \partial_j(\partial_i F \sigma^i_a) \sigma^j_a \right) dt + \partial_i F \sigma^i_a \circ_{0} dW\,,
\end{align}
so that
\begin{align} \label{Average-derivative-F}
    \left\langle \frac{dF}{dt}\right\rangle = \int dx\, P(x) \left( \partial_i F f^i + \left(\frac{1}{2} - \alpha\right) D^{ij} \partial_i \partial_j F + \alpha \partial_j(\partial_i F \sigma^i_a) \sigma^j_a \right)\,.
\end{align}
The average of $\frac{dF}{dt}$ is just the derivative of $\langle F\rangle$ given in (\ref{COnsistency-Prob-Thy}). Taking the derivative of (\ref{COnsistency-Prob-Thy}) we get
\begin{align}
    \left\langle \frac{dF}{dt}\right\rangle = \int dx\, \partial_t P(x,t|x_0,t_0) F(x)\,.
\end{align}
Therefore integrating (\ref{Average-derivative-F}) by parts, we get the Fokker-Planck equation for general regularization
\begin{align}
    \partial_t P & = \left(\frac{1}{2}-\alpha\right)\partial_i \partial_j (D^{ij} P )+ \alpha \partial_i(\sigma^i_a \partial_j (\sigma^j_a P )) - \partial_i(f^i P) \,.
\end{align}
For our special stochastic system with non-multiplicative Gaussian white noise (\ref{Langevin-General}), we have $D^{ij} = \sigma \delta^{ij}$. The Fokker-Planck equation then becomes
\begin{align} \label{Fokker-Planck}
        \partial_t P & = \partial_i \left[\partial^i\left(\frac{\sigma}{2} P \right) - f^i P \right]\,.
\end{align}
This equation can be written $\partial_t P = L P$ where the Fokker-Planck operator $L$ takes the form,
\begin{align} \label{Fokker-Planck-Hamiltonian}
    L = \partial_i \left(\frac{\sigma}{2}\partial^i - f^i  \right) = - \left(\frac{\hat\phi_i^2}{2} \sigma +i \hat \phi_i f^i\right) = - H_{FP}\,,
\end{align}
with $\hat \phi_{i} = -i\partial_i$ with $H_{FP}$ given by (\ref{FP-Hamiltonian}). We then see that the conditional probability is given by
\begin{align}
    P(x,T|x_0,0) = \langle x|e^{TL}|x_0\rangle\,.
\end{align}

\subsection{Ito and Stratonovich in the path integral} \label{Ito-Str-Path-int}

We now want to present the stochastic model in terms of a path-integral and contrast the choice of regulator. The following passages are largely inspired by the work of Ezawa and Klauder~\cite{ezawa_fermions_1985}. We will restrict to a flat target manifold and white noise giving an SDE, 
\begin{align} \label{flat-sde}
    \dot x^i = f^i + \xi^i\,.
\end{align}
As we discussed around (\ref{MSR-path-integral}), the path integral representation of this theory is
\begin{align} \label{PT-Reg-Section}
    Z & = \int D x^i \det \left(\frac{\delta E^i}{\delta x^j}\right) \exp \left(- \int  dt \left[ \frac{(\dot x^i - f^i)^2}{2}\right]\right)\,.
\end{align}
The choice of regulator affects two parts of this expression, 
\begin{align} \label{Import-Terms-Reg-PI}
    \det \left(\frac{\delta E^i}{\delta x^j}\right)\, \qquad \text{and} \qquad \int dt\, \frac{dx_i}{dt} \circ_{\alpha} f^i(x(t))\,.
\end{align}
First, following \cite{ezawa_fermions_1985}, discretize the SDE
\begin{align}
    E^i_n = \frac{x^i_n - x^i_{n-1}}{\epsilon} - \alpha f^i_n - (1-\alpha) f^i_{n-1}\,.
\end{align}
This prescription where $f_n = f(x_n)$ is not the same as but compatible with the discretization used in subsection \ref{App-regularization}.\footnote{We claim that the sums,
\begin{align}
    S^N_A = \sum_{n=1}^N G(x_{\alpha,n}) \Delta W_n\,, \qquad S^N_B = \sum_{n=1}^N \big( \alpha G(x_n) + (1-\alpha) G(x_{n-1}) \big) \Delta W_n\,,
\end{align} agree in the $N\rightarrow\infty$ limit for a function $G(x)$. To see this note that $x_{\alpha,n} = x_{n-1} + \alpha \Delta x_n$ and
\begin{align}
    G(x_{\alpha,n}) & \approx G_{n-1} + \alpha G'_{n-1} \Delta x_n + \frac{\alpha^2}{2} G''_{n-1} (\Delta x_n)^2 + \ldots\,, \\
    G(x_n) & \approx G_{n-1} + G'_{n-1} \Delta x_n + \frac{1}{2} G''_{n-1} (\Delta x_n)^2 + \ldots\,,
\end{align}
and thus
\begin{align}
    S_A^N - S_B^N = \sum_{n=1}^N \frac{\alpha(\alpha-1)}{2} G''_{n-1} (\Delta x_n)^2 \Delta W_n + \mathcal{O}\left((\Delta x_n)^3 \Delta W_n\right)\,.
\end{align}
Since $(\Delta x_n)^2 \sim \Delta t_n$ and $\Delta W_n \sim \sqrt{\Delta t_n}$, the difference of $S^A_N$ and $S^B_N$ is of order $\mathcal{O}\left( (\Delta t_n)^{3/2} \right)$ and is negligible in the mean square limit.}  Our next task is to calculate the determinant of the matrix:
\begin{align} \label{OpenInterval-Disc}
    & \frac{\partial E^i_m}{\partial x^j_n} = \frac{\delta^i_{j}}{\epsilon}\big(\delta_{m,n} - \delta_{m-1,n} \big) - \alpha \delta_{m,n} \partial_j f^i_m - (1-\alpha) \delta_{m-1,n} \partial_j f^i_{m-1}\,, \\ \nonumber
    &= \left(\begin{smallmatrix}
        \frac{\delta^i_{j}}{\epsilon} - \alpha \partial_j f^i_1 & 0 & 0 & \ldots & 0 & 0 \\
        -\frac{\delta^i_{j}}{\epsilon} - (1-\alpha) \partial_j f^i_1 & \frac{\delta^i_{j}}{\epsilon} - \alpha \partial_j f^i_2 & 0 & \ldots & 0 & 0 \\
        0 & -\frac{\delta^i_{j}}{\epsilon} - (1-\alpha) \partial_j f^i_2 & \frac{\delta^i_{j}}{\epsilon} - \alpha \partial_j f^i
        _3 & \ldots & 0 & 0 \\
        \vdots & & & \ddots & & \vdots \\ 0 & 0 & 0 & \ldots & \frac{\delta^i_{j}}{\epsilon} - \alpha \partial_j f^i_{N-1} & 0 \\
        0 & 0 & 0 & \ldots & -\frac{\delta^i_{j}}{\epsilon} - (1-\alpha) \partial_j f^i_{N-1} & \frac{\delta^i_{j}}{\epsilon} - \alpha \partial_j f^i_N
    \end{smallmatrix}\right)\,.
\end{align}
Since this is an upper diagonal matrix its determinant can be trivially calculated giving,
\begin{align}
\begin{aligned}
    \det \left( \frac{\partial E^i_m}{\partial x^j_n} \right) & = \frac{1}{\epsilon^N} \times \prod_{n=1}^N \operatorname{det}\big( \delta^i_{j} - \epsilon \alpha \partial_j f^i \big) = \operatorname{det}  \exp{ \left(- \alpha \int_0^T dt\, \partial_j f^i \right)} \,, \\
    & = \exp{\left( - \alpha \operatorname{Tr} \int_0^T dt\, \partial_j f^i \right)} = \exp{\left(-\alpha \int_0^T dt\, \partial_i f^i \right)}\,.
\end{aligned}
\end{align}
here the product $\prod_{i=1}^N \frac{1}{\epsilon}$ is just the functional determinant $\det(\partial_t)$ which we absorb in the measure. Now we need to investigate the second term in (\ref{Import-Terms-Reg-PI}) which came from expanding the square in the exponent of (\ref{PT-Reg-Section}). We would like to rewrite this in the Stratonovich regularization since it sits inside the path integral where we would like to use standard rules for calculus. We rewrite the second term in (\ref{Import-Terms-Reg-PI}) as
\begin{align} \label{Reg-Int-PI-desc}
\begin{aligned}
   & \int dt\, \frac{dx_i}{dt} \circ_\alpha f^i\big( x(t) \big)  = \sum_{n=0}^N \left( \alpha f^i_n + (1-\alpha) f^i_{n-1} \right) \left( x^i_n - x^i_{n-1} \right) \\
    & = \hlf\sum_{n=0}^N \left(f^i_n + f^i_{n-1}\right)\left( x^i_n - x^i_{n-1} \right) + \left( \alpha - \frac{1}{2} \right) \sum_{n=1}^N \left( f^i_n - f^i_{n-1} \right) \left( x^i_n - x^i_{n-1} \right)\,.
\end{aligned}
\end{align}
The first term is the definition of the Stratonovich integral,
\begin{align}
    \int dt\, \frac{dx_i}{dt} \circ_{1/2} f^i\big( x(t) \big) = \sum_{n=0}^N \frac{f^i_n + f^i_{n-1}}{2}\left( x^i_n - x^i_{n-1} \right)\,.
\end{align}
We Taylor expand the second term,
\begin{align}
    f^i_n-f^i_{n-1}=\partial_j f^i\left(x_{n-1}\right) \Delta x^j_n+\mathcal{O}\left((\Delta x)^2\right)\,,
\end{align}
so the last term on the right-hand side of (\ref{Reg-Int-PI-desc}) becomes,
\begin{align}
    \left(\alpha-\frac{1}{2}\right) \sum_{n=1}^N \partial_j f^i\left(x_{n-1}\right) \Delta x^j_n \Delta x^i_n+\mathcal{O}\left((\Delta x)^{3}\right)\,.
\end{align}
Since $\Delta x^i_n$ is a Wiener process $\Delta x^i_n \Delta x^{j}_{n} \rightarrow \delta_{ij} \Delta t$ thus
\begin{align} \label{alpha-regularized-integral}
    \int dt\, \frac{dx_i}{dt} \circ_\alpha f^i\big( x(t) \big) = \int dt \, \dot x_i f^i - \left(\frac{1}{2} - \alpha \right) \int dt \, \partial_i f^i\,,
\end{align}
where in the right hand side we are using the Stratonovich product but we are dropping $\circ_{1/2}$ from now on. We therefore get
\begin{align} \label{alpha-fam-PI}
\begin{aligned}
    &\int Dx^i \det\left( \frac{\delta E^i}{\delta x^j} \right) \exp{\left(-\frac{1}{2} \int dt \left(\dot x^i - f^i\right)^2\right)}\,, \\
    & = \int Dx^i \exp{\left(-\alpha \int dt\, \partial_i f^i \right)} \exp{\left(- \int dt \left(\frac{\dot x_i^2}{2}+ \frac{f_i^2}{2}\right) + \int dt \,\dot x_i \circ_\alpha f^i \right)} \,,\\
    & = \int Dx^i \exp{\left(-\alpha \int dt\, \partial_i f^i \right)} \exp{\left(- \int dt \left(\frac{\dot x_i^2}{2}+ \frac{f_i^2}{2} - \,\dot x_i f^i +\frac{\partial_i f^i}{2} - \alpha \partial_i f^i \right) \right)} \,,\\
    & = \int Dx^i \exp{\left(- \int dt \left(\frac{\dot x_i^2}{2}+ \frac{f_i^2}{2} - \,\dot x_i f^i +\frac{\partial_i f^i}{2} \right) \right)}\,,
\end{aligned}
\end{align}
So physical quantities computed via the path-integral for this system with white noise are independent of the choice of regularization. 

\section{Finite-dimensional Grassmann Integrals} \label{App-Grassmanian-Integrals}

\subsection{Representing determinants as real or complex Grassmann integrals.}  \label{App-Real-vs-Complex}

Consider an $n\times n$ matrix $E$. We want to prove that the identity
\begin{align} \label{Det-Id}
    \det(E) = \int \prod_{i=1}^n d\eta_i d\tilde \eta_i \exp \sum_{ij} \tilde \eta_i E_{ij} \eta_j\,,
\end{align}
is independent on whether we assume that $\eta_i,\tilde\eta_i$ are real or complex Grassmann variables. The proof hinges on the fact that if we assume these two variables are real, independent Grassmann variables then
\begin{align}
    \int d \eta d\tilde \eta \, \tilde \eta \eta =1\,.
\end{align}
Conversely assume they are complex given by
\begin{align}
    &\eta = \frac{\eta^x+i\eta^y}{\sqrt{2}}\,, \qquad \tilde\eta = \frac{\eta^x-i\eta^y}{\sqrt{2}}\,,
    &\begin{pmatrix}
        \eta \\ \tilde \eta
    \end{pmatrix} = \begin{pmatrix}
        \frac{1}{\sqrt{2}} & \frac{i}{\sqrt{2}} \\
        \frac{1}{\sqrt{2}} & -\frac{i}{\sqrt{2}}
    \end{pmatrix} \begin{pmatrix}
        \eta^x \\ \eta^y
    \end{pmatrix}\,,
\end{align}
where $\eta^x$ and $\eta^y$ are real and thus satisfy $\int d\eta^x \eta^x = \int d\eta^y \eta^y =1$. Then
\begin{align}
\begin{aligned}
    &d\eta d\tilde \eta = \left[\det\begin{pmatrix}
        \frac{1}{\sqrt{2}} & \frac{i}{\sqrt{2}} \\
        \frac{1}{\sqrt{2}} & -\frac{i}{\sqrt{2}}
    \end{pmatrix}\right]^{-1} d\eta^ x d\eta^y = i d\eta^ x d\eta^y\,, \\
    &\tilde \eta \eta = \frac{1}{2}(\eta^x-i\eta^y)(\eta^x+i\eta^y) = \frac{i}{2}(\eta^x \eta^y - \eta^y \eta^x) =-i \eta^y \eta^x\,.
\end{aligned}
\end{align}
Therefore under this assumption
\begin{align}
    \int d \eta d\tilde \eta \, \tilde \eta \eta = \int d\eta^ x d \eta^y \, \eta^y \eta^x = 1\,.
\end{align}
Now we go back to (\ref{Det-Id}). This identity can be written as
\begin{align} \label{Det-Id-Ex}
    \begin{aligned}
    \int \prod_{k=1}^n d\eta_k d\tilde \eta_k \exp \sum_{ij} \tilde \eta_i E_{ij} \eta_j & = \int \left(\prod_{k=1}^n d\eta_k d\tilde \eta_k\right) \prod_{i=1}^n \exp \left( \tilde\eta_i \sum_{j} \tilde E_{ij} \eta_j \right) \,, \\
    & = \int \left(\prod_{k=1}^n d\eta_k d\tilde \eta_k\right) \prod_{i=1}^n\left(1+  \tilde\eta_i \sum_{j} \tilde E_{ij} \eta_j\right) \,, \\
    & = \int \left(\prod_{k=1}^n d\eta_k d\tilde \eta_k\right) \prod_{i=1}^n  \tilde\eta_i \sum_{j} \tilde E_{ij} \eta_j\,.
    \end{aligned}
\end{align}
Now notice that
\begin{align}
\begin{aligned}
    \prod_{i=1}^n  \tilde\eta_i \sum_{j} E_{ij} \eta_j & = \tilde\eta_1 \left(\sum_{j_1} E_{1j_1} \eta_{j_1}\right) \tilde\eta_2 \left(\sum_{j_2} E_{2j_2} \eta_{j_2}\right) \ldots \tilde\eta_n \left(\sum_{j_n} E_{nj_n} \eta_{j_n}\right) \,, \\
    & = \sum_{j_1\ldots j_n} \tilde \eta_1 \eta_{j_1} \tilde \eta_1 \eta_{j_2} \ldots \tilde \eta_n \eta_{j_n} \, E_{1j_1} E_{2j_2} \ldots E_{nj_n} \,,\\
    & = \tilde \eta_1 \eta_1 \ldots \tilde \eta_n \eta_n \left( \sum_{j_1\ldots j_n} \text{sgn}(j_1,\ldots ,j_n) E_{1j_1} E_{2j_2} \ldots E_{nj_n} \right) \,, \\
    &= \left(\prod_{i=1}^n \tilde \eta_i \eta_i \right) \det(E)\,.
\end{aligned}
\end{align}
Plugging this result back into (\ref{Det-Id-Ex}) we get
\begin{align}
    \det(E) \int\left(\prod_{k=1}^n d\eta_k d\tilde \eta_k\right) \left(\prod_{i=1}^n \tilde \eta_i \eta_i \right) = \det(E) \prod_{i=1}^n \int d\eta_i d\tilde \eta_i \, \tilde \eta_i \eta_i = \det(E)\,,
\end{align}
since
\begin{align}
    \prod_{i=1}^n \int d\eta_i d\tilde \eta_i \, \tilde \eta_i \eta_i = 1\,,
\end{align}
which is independent of whether we assume that these are real or complex Grassmann variables.

\subsubsection*{\ul{\it The usual relation between Pfaffians and determinants}}

Before we move on and give a proof of (\ref{det(B)-New}) for finite-dimensional integrals, we want to briefly review the standard relation between Pfaffians and determinants: 
\begin{align} \label{standardsign}
    \text{Pf}\begin{pmatrix}
        0 & E \\
        - E^T & 0
    \end{pmatrix} = (-1)^{\frac{n(n-1)}{2}} \det(E)\,.
\end{align}
By definition the Pfaffian is given by
\begin{align}
    \text{Pf}\begin{pmatrix}
        0 & E \\
        - E^T & 0
    \end{pmatrix} & = \int d\theta_{2n} \ldots d\theta_1 \exp{ \left[ \frac{1}{2} \begin{pmatrix}
        \theta_i & \theta_{i+n}
    \end{pmatrix}
    \begin{pmatrix}
        0 & E_{ij} \\
        - E_{ji} & 0
    \end{pmatrix} \begin{pmatrix}
        \theta_j \\ \theta_{j+n}
    \end{pmatrix}\right]} \,, \\
    & = \int d\theta_{2n} \ldots d\theta_1 \exp{\sum_{i,j=1}^n \theta_i E_{ij} \theta_{j+n}}\,.
\end{align}
Notice the convention chosen for the integration measure: $d\theta_{2n} \ldots d\theta_1$. This choice gives the usual phase appearing in \C{standardsign}; see, for example, Section 1.7.2 of \cite{Zinn-Justin2021qftc}. If we instead choose $d\theta_{1} \ldots d\theta_{2n}$ then we would get $(-1)^{\frac{n(n+1)}{2}}$ for the phase factor multiplying $\det(E)$.

Now define $\tilde\eta_i = \theta_i $ and $\eta_i = \theta_{i +n}$ with $i=1,\ldots,n$ and write
\begin{align}
\begin{aligned}
    \int d\theta_{2n} \ldots d\theta_1 \exp{\sum_{ij=1}^n \theta_i E_{ij} \theta_{j+n}} 
    & = \int d\eta_n ... d\eta_1 d \tilde\eta_n ... d\tilde\eta_{1} \exp{\sum_{ij=1}^n \tilde\eta_i E_{ij} \eta_{j}} \,, \\
    & = \int d\eta_n ... d\eta_1 d \tilde\eta_n ... d\tilde\eta_{1}\, \tilde \eta_1 \eta_1 \ldots \tilde \eta_n \eta_n \det (E)\,.
\end{aligned}
\end{align}
To perform the Grassmann integral and get the correct sign factor we reorganize the integration measure as follows:
\begin{align}
\begin{aligned}
    d\eta_n ... d\eta_1 d \tilde\eta_n ... d\tilde\eta_{1} & = (-1)^{\sum_{a=1}^n(n-a)} \prod_{i=1}^n d \eta_i d \tilde\eta_i \\
    &= (-1)^{\frac{n(n-1)}{2}} \prod_{i=1}^n d \eta_i d \tilde \eta_i\,.
\end{aligned}
\end{align}
Plugging this back gives,
\begin{align}
\begin{aligned}
    \text{Pf}\begin{pmatrix}
        0 & E \\
        - E^T & 0
    \end{pmatrix} & = (-1)^{\frac{n(n-1)}{2}} \det(E) \int \prod_{i=1}^n d \eta_i d\tilde \eta_i \, \tilde \eta_1 \eta_1 \ldots \tilde \eta_n \eta_n \,, \\
    & = (-1)^{\frac{n(n-1)}{2}} \det(E)\,,
\end{aligned}
\end{align}
which is the desired identity.

\subsection{The Pfaffian identity for finite-dimensional integrals.} \label{App-FermionicProofPffafian}

The new $N=1$ supersymmetry intrinsic to nonreciprocal stochastic theories relies on the identity
\begin{align} \label{Pfaffian-New-eq}
    \text{Pf}\begin{pmatrix}
        M & E \\
        - E^T & 0
    \end{pmatrix} = (-1)^{\frac{n(n-1)}{2}} \det(E)\,,
\end{align}
where $M$ and $E$ are $n\times n$ matrices. Here we want to prove this identity for finite-dimensional matrices, while in Appendix \ref{timediscrete} we extend the proof to infinite-dimensional matrices. To prove this, we use the Grassmann formulation of the Pfaffian. 

By definition the Pfaffian is given by,
\begin{align}
    \text{Pf}\begin{pmatrix}
        M & E \\
        - E^T & 0
    \end{pmatrix} & = \int d\theta_{2n} \ldots d\theta_1 \exp{\frac{1}{2} \sum_{i,j=1}^n \begin{pmatrix}
        \theta_i & \theta_{i+n}
    \end{pmatrix}\begin{pmatrix}
        M_{ij} & E_{ij} \\
        - E_{ji} & 0
    \end{pmatrix} \begin{pmatrix}
        \theta_j \\ \theta_{j+n}
    \end{pmatrix}}\,, \\
    & = \int d\theta_{2n} \ldots d\theta_1 \exp{\sum_{i,j=1}^n \left(\frac{1}{2} \theta_i \theta_j M_{ij} + \theta_i E_{ij} \theta_{j+n} \right)}\,.
\end{align}
At this point we can basically see by inspection that to get the requisite number of $\theta_{j+n}$ fermions for a volume form, no terms from $M$ can participate. Since this is a key point, let us be more explicit. Again define $\tilde\eta_i = \theta_i $ and $\eta_i = \theta_{i +n}$ for $i=1,\ldots,n$ and write
\begin{align}
\begin{aligned}
    &\int d\theta_{2n} \ldots d\theta_1 \exp{\sum_{ij=1}^n \left(  \frac{1}{2} \theta_i \theta_j M_{ij} + \theta_i E_{ij} \theta_{j+n} \right)} \\
    & = \int d\eta_n ... d\eta_1 d \tilde\eta_n ... d\tilde\eta_{1} \exp{\sum_{ij=1}^n \left(  \frac{1}{2} \tilde\eta_i \tilde\eta_j M_{ij} + \tilde\eta_i E_{ij} \eta_{j} \right)}\,, \\
    &= \det(E) \int d\eta_n' ... d\eta_1' d \tilde\eta_n ... d\tilde\eta_{1} \exp{\sum_{ij=1}^n \left(  \frac{1}{2} \tilde\eta_i \tilde\eta_j M_{ij} + \tilde\eta_i \eta_{i}' \right)}\,,
\end{aligned}
\end{align}
In the last line we made the substitution
\begin{align}
    \eta'_i = E_{ij} \eta_j\,,
\end{align}
which gives
\begin{align}
    d\eta_n ... d\eta_1 = \det(E)\, d\eta_n' ... d\eta_1'\,.
\end{align}
The leftover integral is
\begin{align}
\begin{aligned}
    &\int d\eta_n' ... d\eta_n1' d \tilde\eta_n ... d\tilde\eta_{1} \exp{ \left(  \frac{1}{2} \sum_{ij=1}^n \tilde\eta_i \tilde\eta_j M_{ij} + \sum_{i=1}^n\tilde\eta_i \eta_{i}' \right)} \\
    & = \int d\eta_n' ... d\eta_1' d \tilde\eta_n ... d\tilde\eta_{1} \prod_{i=1}^n \exp{\tilde\eta_i \left(  \frac{1}{2} \sum_{j=1}^n \tilde\eta_j M_{ij} + \eta_{i}' \right)}\,, \\
    &  = \int d\eta_n' ... d\eta_1' d \tilde\eta_n ... d\tilde\eta_{1} \prod_{i=1}^n\left[ 1 + \tilde\eta_i \left(  \frac{1}{2} \sum_{j=1}^n \tilde\eta_j M_{ij} + \eta_{i}' \right)\right]\,.
\end{aligned}
\end{align}
For this integral to be non-zero we must have one and only one pair $\tilde \eta_i \eta_i'$ for every $i$. This means we cannot take any $\tilde \eta_i $ from the $M$ coupling which gives, 
\begin{align}
\begin{aligned}
    \text{Pf}\begin{pmatrix}
        M & E \\
        - E^T & 0
    \end{pmatrix} & = (-1)^{\frac{n(n-1)}{2}} \det(E) \int \prod_{i=1}^n d \eta_i' d\tilde \eta_i \, \tilde \eta_1 \eta_1' \ldots \tilde \eta_n \eta_n' \,,\\
    & = (-1)^{\frac{n(n-1)}{2}} \det(E)\,,
\end{aligned}
\end{align}
which is the desired identity.

\subsection{Time discretization of the path-integral} \label{timediscrete}

In Section \ref{MSR} we proposed that the fermion terms in the path integral (\ref{N=1-SUSY-QM}), which take the form
\begin{align}
    & \int D\chi D\psi \exp \left[- \frac{1}{2} \int dt\, \big(\psi_i \quad \chi^i \big)\begin{pmatrix}
       \partial_t\delta^{ij} & \partial_t\delta^i_j - \partial_jf^i \\
       \partial_t\delta^j_i + \partial_if^j & 0
    \end{pmatrix} \begin{pmatrix}
        \psi_j \\ \chi^j
    \end{pmatrix} \right]\,,
\end{align}
give the functional determinant $\det\big(\partial_t\delta^j_i + \partial_i f^j\big)$. To see this we discretize the various terms in the fermionic action,
\begin{align}
    \psi_i \dot \psi_i & \rightarrow \psi_{in} \frac{\psi_{in} - \psi_{in-1}}{\epsilon} = \psi_{im}\underbrace{\frac{\delta^{ij}}{2\epsilon} \big(\delta_{m,n-1}-\delta_{m-1,n}\big)}_{M_{im,jn}} \psi_{jn}\,, \label{Discretization-Fermion-Act}\\ 
    \psi_i \dot \chi^i - \psi_i \partial_j f^i \chi^j & \rightarrow \psi_{im}\underbrace{\left[ \frac{\delta^i_{j}}{\epsilon}\big(\delta_{m,n} - \delta_{m-1,n} \big) - \alpha \delta_{m,n} \partial_j f^i_m - (1-\alpha) \delta_{m-1,n} \partial_j f^i_{m-1}\right]}_{E_{im,jn}} \chi^j_n\,, \nonumber \\
    -\dot\chi^i \psi_i + \chi^i \partial_i f^j \psi_j & \rightarrow \chi_{im}\underbrace{\left[ \frac{\delta^i_{j}}{\epsilon}\big(-\delta_{m,n} + \delta_{m,n-1} \big) + \alpha \delta_{m,n} \partial_i f^j_m + (1-\alpha) \delta_{m,n-1} \partial_i f^j_{n-1}\right]}_{-E_{jn,im}} \psi^j_n\,, \nonumber
\end{align}
where $n,m=1,\ldots,N$. At the end we send $N\rightarrow\infty$ and $\epsilon \rightarrow 0$ with $N\epsilon \rightarrow T$. The discretization in the first line of (\ref{Discretization-Fermion-Act}) ensures that $M^T = - M$. The second line uses precisely the same discretization and $\alpha$ regularization used in (\ref{OpenInterval-Disc}). After discretization our path integral can be written as,
\begin{align}
    \int \prod_{m,n,i,j} d \psi_{im} d\chi_{im} \exp \left[ -\frac{1}{2} \sum_{m,n,i,j} \epsilon \begin{pmatrix}
        \psi_{im} & \chi_{im}
    \end{pmatrix}\begin{pmatrix}
        M_{im,jn} & E_{im,jn} \\
        -E_{jn, im} & 0
    \end{pmatrix}\begin{pmatrix}
        \psi_{jn} \\
        \chi_{jn}
    \end{pmatrix} \right] \,.
\end{align}
Now the problem becomes a straightforward application of the results from the last section giving, 
\begin{align} \label{App-eq-we-want-proof}
    \int D\chi D\psi \exp \left[- \frac{1}{2} \int dt\, \big(\psi_i \quad \chi^i \big)\begin{pmatrix}
       \partial_t\delta^{ij} & \partial_t\delta^i_j - \partial_jf^i \\
       \partial_t\delta^j_i + \partial_if^j & 0
    \end{pmatrix} \begin{pmatrix}
        \psi_j \\ \chi^j
    \end{pmatrix} \right] = \det\big(\partial_t\delta^j_i - \partial_i f^j\big)\,,
\end{align}
where we have absorbed any constant phase in the choice of measure.

\section{Symplectic Quantization}
\label{symplectic}

In this Appendix we discuss  symplectic quantization of the actions (\ref{MSR-Action}) and (\ref{S-New-No-Auxiliary}). This is a more formal way of quantizing the system is a good way to check that the quantization conditions (\ref{Bosonic-Commutation}) and (\ref{Fermionic-Commutation}) are the correct ones for the action (\ref{S-New-No-Auxiliary}). The  point of symplectic quantization is to extract a symplectic form from the action and use it to define Poisson brackets. We then quantize those Poisson brackets. Using this more robust formalism we can address the issue in footnote \ref{subtle}: even though (\ref{QtildeQ-Coordinates}) tells us that $[\delta,\tdelta] \neq -2 \epsilon \tepsilon \partial_t$ we found that $\{Q,\tilde Q\} = 2H_{MSR}$. How are these two statements compatible?

\subsubsection*{\ul{\it Brief overview of symplectic quantization}}

Let us briefly describe the method of symplectic quantization. We first Wick rotate our Euclidean action to Minkowski signature\footnote{It is possible to obtain the Poincar\'e 1-form and symplectic form in either Euclidean or Minkowski signature. However the canonical definitions are usually given in Minkowski signature; hence to avoid confusion, we choose to Wick rotate.}. The next step is to vary the action keeping track of the total derivative term, 
\begin{align}
    \delta S = \int dt \left[ (\text{e.o.m.})\delta \phi + \frac{d}{dt} \Theta \right]\,.
\end{align}
Here $\Theta$ is the Poincar\'e 1-form, which is linear in $\delta \phi$. We then view $\Theta$ as a $1$-form $p_id\phi^i$ in phase space.  Here the exterior derivative $d$ acts on phase space (not spacetime). The exterior derivative of $\Theta$ is the symplectic form: $\omega = d \Theta$. It is convenient to label the phase space coordinates $(p_i, \phi^j)$ collectively by $z^A$. In a coordinate chart, $\omega = \frac{1}{2} \omega_{AB} dz^A \wedge dz^B$ where we assume $\omega_{AB}$ is invertible with $\omega^{AC} \omega_{CB} = \delta^A_B$. Poisson brackets between two scalar functions $f$ and $g$ are defined as follows,
\begin{align} \label{Poisson-Bracket-omega}
    \{f,g\}_{P} \equiv \omega(X_g,X_f) = \omega^{AB} \, \partial_A f \partial_B g \,.
\end{align}
Here $X_f$ and $X_g$ are Hamiltonian vector fields defined by,
\begin{align} \label{Ham-Vec-Field}
    \left( X_f \right)^A = \omega^{AB} \partial_B f\,.
\end{align}
Finally quantization of the Poisson bracket is the map:
\begin{align} \label{Quantization-Map}
\begin{aligned}
    &\{\,\, ,\, \}_{P} \rightarrow \frac{1}{i}[\,\, ,\,]_{\pm}\,, 
\end{aligned}
\end{align}
where the $\pm$ means commutator if either variable is bosonic or anticommutator if both variables are fermionic.

As warm up examples, we perform symplectic quantization of two simple systems:
\begin{align} \label{Toy-Actions}
    S^b = \int dt\, \frac{\dot \phi^2}{2}\,, \qquad S^f = \int dt\, \tpsi i\dot \psi\,.
\end{align}
The variations of these actions take the form,
\begin{align}
    & \delta S^b = \int dt\left[-\ddot \phi \delta \phi + \frac{d}{dt}\left( \dot \phi \delta \phi \right) \right]\,, \\ &\delta S^f = \int dt\left[ \delta \tpsi i \dot \psi + \delta \psi i \dot\tpsi +  \frac{d}{dt}\left( i\frac{\tpsi}{2}\delta \psi + i \frac{\psi}{2}\delta\tpsi \right) \right]\,.
\end{align}
Using $p=\dot\phi$ the Poincar\'e 1-form and the symplectic form are\footnote{Here we note that $d\psi$ is a boson if $\psi$ is Grassmann.  This implies, for example, that $d\tpsi \wedge d\psi = d\psi \wedge d\tpsi$, or more generally,
\begin{align} \label{Deligne-Convention}
    A \wedge B = \frac{1}{2}\left( A\otimes B -(-1)^{|A|\cdot|B|} B\otimes A \right)\,,
\end{align}
where $|\mathcal{O}|$ is the Grassmann parity of $\mathcal{O}$.}
\begin{align}
\begin{aligned}
    &\theta_b = p d\phi\,, && \qquad \theta_f = i\frac{\tpsi}{2}d\psi + i \frac{\psi}{2}d\tpsi\,, \\
    &\omega_b = dp\wedge d\phi\,, && \qquad \omega_f = i d\tpsi \wedge d\psi\,.
\end{aligned}
\end{align}
From these symplectic forms, we obtain the Poisson brackets
\begin{align}
    \{\phi,p\}_P = \omega(X_{p},X_{\phi}) = 1\,, \qquad \{\psi,\tpsi\}_P = \omega(X_{\tpsi},X_{\psi}) = -i\,. \nonumber
\end{align}
Finally using (\ref{Quantization-Map}) we quantize to get
\begin{align}
    [\phi,p] = i\,, \qquad \{\psi,\tpsi\} = 1\,,
\end{align}
which are the standard quantization conditions for (\ref{Toy-Actions}).

\subsubsection*{\ul{\it Symplectic quantization of the SUSY action (\ref{S-New-No-Auxiliary})}}

We start by Wick rotating the action,
\begin{align} \label{MSR-Action-with-A}
    S_{SUSY} & = \int dt \left( \frac{\dot \phi_i^2}{2} + i \dot \phi^i A_i - \frac{f_i^2}{2} +\frac{1}{2} \psi^i i \dot \psi_i + \chi^i i \dot \psi_i + \chi^j \partial_jf^i  \psi_i \right)\,.
\end{align}
The variation of this action is
\begin{align}
    \delta S & = \int dt \left[\ldots + \frac{d}{dt}\left( (\dot \phi_i + i A_i) \delta \phi^i + \frac{i}{2}(\psi^i + \chi^i) \delta \psi_i + \frac{i}{2} \psi_i \delta \chi^i \right) \right]\,.
\end{align}
From the total derivative term we read off the Poincar\'e 1-form,
\begin{align}
    \Theta = (\pi_i + i A_i) d\phi^i + \frac{i}{2}(\psi_i + \chi_i) d \psi^i + \frac{i}{2} \psi^i d\chi_i\,,
\end{align}
where we used $\pi_i = \dot\phi_i$. The symplectic form is
\begin{align}
    \omega = d\pi_i \wedge d \phi^i - i \frac{\mathcal{F}_{ij}}{2} d\phi^i \wedge d\phi^j + \frac{i}{2} d\psi_i \wedge d\psi^i + i d\chi_i \wedge d\psi^i\,.
\end{align}
Using (\ref{Poisson-Bracket-omega}) we get
\begin{align} \label{Sympletic-Boson-Pois}
    &\{\phi^i,\pi_j\}_{P}  = \delta^i_{j}\,, \qquad \{\pi_i,\pi_j\}_{P} = - i \mathcal{F}_{ij} \\
    &\{\psi_i,\psi_j\}_{P} = 0\,, \qquad \{\psi_i,\chi^j\}_{P} = - i\delta^j_{i}\,, \qquad \{\chi^i,\chi^j\}_{P} = i \delta^{ij}\,.
\end{align}
Finally using (\ref{Quantization-Map}) we get
\begin{align} \label{Sympletic-Boson-Comm}
    &[\phi^i,\pi_j] = i\delta^i_{j}\,, \qquad [\pi_i,\pi_j] = \mathcal{F}_{ij}\,, \\
    &\{\psi_i,\psi_j \} = 0\,, \qquad \{ \chi^i, \psi_j \} = \delta^i_{j}\,, \qquad \{\chi_i, \chi_j \} = -\delta_{ij}\,,
\end{align}
which are exactly the same as (\ref{Bosonic-Commutation}) and (\ref{Fermionic-Commutation}) obtained using (\ref{Standard-Comm-Rel}).

\subsubsection*{\ul{\it Symplectic quantization of the MSR action (\ref{MSR-Action})}}

The Wick rotated MSR action is
\begin{align}
    S_{MSR} & = \int dt \left( \frac{\dot\phi_a^2}{2} + i \dot\phi_a A_a - \frac{f_a^2}{2} + \tilde \psi_a i\dot \psi_a + \tilde \psi_b \partial_b f_a \psi_a \right)\,.
\end{align}
The bosonic part of this action is exactly the same as before. The variation of the fermionic part is
\begin{align}
    \delta S^{F}_{MSR} & = \int dt \left[ \ldots + \frac{d}{dt}\left( \frac{i}{2}\tilde \psi^i \delta \psi_i + \frac{i}{2} \psi_i \delta \tpsi^i \right) \right]\,.
\end{align}
Thus the Poincar\'e 1-form and the symplectic form for the fermions are, 
\begin{align}
    \theta^F_{MSR}  = \frac{i}{2}\tilde \psi_i d \psi^i + \frac{i}{2} \psi_i d \tpsi^i\,, \qquad
    \omega^F_{MSR} = i d\tpsi^i \wedge d\psi_i\,,
\end{align}
with the full symplectic form:
\begin{align}
    \omega & = d\pi_i \wedge d\phi^i - i \frac{F_{ij}}{2} d\phi^i \wedge d\phi^j + i d\tpsi^i \wedge d\psi_i\,. 
\end{align}
The Poisson bracket and quantum commutation relations for the bosonic degrees of freedom will be exactly the same as (\ref{Sympletic-Boson-Pois}) and (\ref{Sympletic-Boson-Comm}). For the fermions we now have,
\begin{align}
\begin{aligned}
    &\{ \tilde\psi^i, \psi_j \}_P = -i\delta^i_{j}\,, && \qquad \{\tilde\psi^i, \tilde\psi^j \}_P = 0\,, && \qquad \{\psi_i, \psi_j \}_P = 0\,, \\
    &\{ \tilde\psi^i, \psi_j \} = \delta^i_{j}\,, && \qquad \{\tilde\psi^i, \tilde\psi^j \} = 0\,, && \qquad \{\psi_i, \psi_j \} = 0\,,
\end{aligned}
\end{align}
which agree with (\ref{Comm-Rel-Fermions-MSR}).

\subsubsection*{\ul{\it How is $\{Q,\tilde Q\} = 2H_{MSR}$ but  $[\delta,\tdelta] \neq -2 \epsilon \tepsilon \partial_t$?}}

To address this question we first note that a linear transformation of the fields should be understood as a vector acting on the phase space. Specifically the transformations (\ref{Q-In-MSR}) and (\ref{Qtilde-In-MSR}) act as vectors,
\begin{align}
    \delta & = \epsilon \Big( -i \partial^j f^i \tpsi_j \,\partial_{\pi_i} - \tpsi^i \,\partial_{\phi_i} + (-i\pi^i - f^i)\partial_{\psi_i} \Big) \,,
    \\
    \tdelta & = \tepsilon \Big( - i \partial^i f^j \psi_j \,\partial_{\pi_i} + \psi^i \,\partial_{\phi_i} + (i\pi^i - f^i) \partial_{\tpsi_i} \Big)\,,
\end{align}
acting on phase space with coordinates $(\pi_i,\phi^i,\tpsi^i,\psi_i)$.

In symplectic mechanics a Hamiltonian vector field, $V$, is a vector field which can be implemented by a scalar function $G_V$ via Poisson bracket:
\begin{align} \label{Ham-Vec-Field-cond}
    V \mathcal{O} = \{G_V,\mathcal{O}\}_P\,.
\end{align} 
Said differently, a vector field $V$ is Hamiltonian if and only if there exists a function $G$ such that
\begin{align} \label{Ham-Vec-Field-Cond}
    \iota_V \omega = d G_V\,.
\end{align}
The symplectic form is closed, $d\omega=0$, so the Lie derivative $\mathcal{L}_V \omega$ simplifies:
\begin{align}
    \mathcal{L}_V \omega = d(\iota_V \omega)\,.
\end{align}
So if $d(\iota_V \omega) = 0$ then the symplectic form is conserved along the flow generated by $V$; in this case $V$ is a symplectomorphism. All Hamiltonian vector fields are symplectomorphisms but not all symplectomorphisms are Hamiltonian vector fields. 

The operators $Q$ and $\tilde Q$ are given by
\begin{align} 
    & Q =- \tpsi^i \big( i\pi_i + f_a \big)\,, \qquad \tilde Q = \psi^i\big( i\pi_i - f_i \big)\,.
\end{align}
It is easy to check that
\begin{align}
    \iota_{\delta}  \omega = d\big(\epsilon iQ\big)\,,
\end{align}
and so $\delta$ can be implemented using the supercharge $Q$ via $\delta \mathcal{O} = \{i\epsilon Q, \mathcal{O}\}_{P}$ in the classical theory and $\delta \mathcal{O} = [\epsilon Q, \mathcal{O}]$ in the quantum theory. The second transformation is more subtle. We first notice that
\begin{align}
    \iota_{\tdelta} \omega = \tepsilon \Big( -\psi^i d\pi_i - i \partial^j f^i \psi_j d\phi_i + i(i\pi^i - f^i)d\psi_i \Big) \neq d\big(\tepsilon i \tilde Q\big)\,,
\end{align}
so $\tdelta$ cannot be implemented by $\tilde Q$. However, the problem with this transformation is even worse because
\begin{align} \label{symplectomorphism-condition}
    d\left(\iota_{\tdelta} \omega\right) = - \tepsilon i \left( d\psi^i \wedge \phi^j + d\phi^i \wedge d\phi^j \frac{\psi^k \partial_k}{2}\right) F_{ij}\,,
\end{align}
so $\iota_{\tdelta} \omega$ is not even a closed form. This means $\tdelta$ fails to be a symplectomorphism. Not only does $\tilde Q$ not implement $\tdelta$, there is no local scalar function which could possibly implement $\tdelta$ via a Poisson bracket.

This raises the question: if the operator $\tilde Q$ obtained from (\ref{Breaking-PS}) does not implement $\tdelta$ then what transformation $\tdelta'$ does it generate?
The Hamiltonian vector field obtained from $\tilde Q$ using \C{Ham-Vec-Field} is
\begin{align} \label{tildeQ-Ham-vec-field}
    \tdelta' = X_{\tepsilon i \tilde Q} & = \tepsilon \Big( -i(\partial_i f_j + \mathcal{F}_{ij}) \psi_j\,\partial_{\pi_i} + \psi_i\, \partial_{\phi_i} + (i\pi_i - f_i )\partial_{\tpsi_i}  \Big)\,.
\end{align}
Notice that $\tdelta$ and $\tdelta'$ differ only by the term $\mathcal{F}_{ij} $ that appears in the direction $\partial_{\pi_i}$. To be more explicit these two transformations only differ in the way they act on $\dot\phi^i$,
\begin{align} \label{hat-tdelta-dotphi}
    \tdelta \dot\phi^i = \tepsilon \dot \psi^i \,,\qquad \tdelta' \dot \phi^i = \tepsilon \left(\dot \psi^i -  \mathcal{F}^{ij} \psi_j \right)\,.
\end{align}
The extra term in $\tdelta'\dot\phi^i$ is exactly what we need to cancel the spurious term in (\ref{QtildeQ-Coordinates}). If we use $\tdelta'$ instead of $\tdelta$ we get
\begin{align}
    \begin{aligned}
    & [\,\delta , \tdelta' \,] \phi^i = \epsilon \tilde \epsilon \left( - 2 \dot \phi^i \right), \\
    &[\,\delta , \tdelta' \,] \psi^i = \epsilon \tilde \epsilon \left(-2 \dot \psi^i \right), \\ & [\,\delta , \tdelta' \,] \tilde \psi^i = \epsilon\tilde \epsilon \left( - 2 \dot{\tilde{\psi}}^i\right)\,,    \end{aligned}
\end{align}
showing $[\delta,\tdelta'] = \epsilon \tepsilon(-2\partial_t)$. Since $[\,\delta,\tdelta' \,] \mathcal{O} = -\epsilon \tepsilon \big(  \big[\{Q,\tilde Q\},\mathcal{O}\big] \big)$ and the Euclidean Heisenberg equations of motion are $\partial_t \mathcal{O} =[H,\mathcal{O}]$ we expect that ${\{Q,\tilde Q\} = 2 H_{MSR}}$. Indeed it is easy to check that
\begin{align} \label{Hamiltonian}
    \{ Q ,\tilde Q \} 
    = & 2 \left(  \frac{1}{2}(p_i -i A_i)^2 + \frac{1}{2}(\partial_i V + A_i)^2 - (\partial_j \partial_i V + \partial_i A_j)\frac{[\tilde \psi_i, \psi_j]}{2} \right)\,.
\end{align}
Since these two fermionic operators anticommute to $H$, one might hope that if we use $\tdelta'$ instead of $\tdelta$ then maybe the system would be supersymmetric. This hope, however, is frustrated because 
it is not a symmetry of the system. This can be seen by recalling that for general operators $G$ and $\mathcal{O}$, 
\begin{align}
\begin{aligned}
    [G,\partial_t{\mathcal{O}}] & = [G,[H,\mathcal{O}]] = [H,[G,\mathcal{O}]] + [\mathcal{O},[H,G]]\,, \\
    & = \partial_t[G,\mathcal{O}] + [\mathcal{O},[H,G]]\,.
\end{aligned}
\end{align}
So $[G,\partial_t{\mathcal{O}}] = \partial_t [G,\mathcal{O}]$ if and only if $G$ is a symmetry of the Hamiltonian with ${[G,H] = 0}$. Applying this identity to (\ref{hat-tdelta-dotphi}) we get
\begin{align}
    \underbrace{[\tepsilon\tilde Q, \dot \phi^i]}_{\tdelta' \dot\phi^i} = \underbrace{\partial_t [\tepsilon\tilde Q,\phi^i]}_{\partial_t (\tepsilon \psi^i)} + \underbrace{[\phi^i,[H_{MSR},\tepsilon\tilde Q]]}_{-\tepsilon \mathcal{F}^{ij}\psi_j}\,,
\end{align}
from which we see that $[\tilde Q,H_{MSR}] \neq 0$. Therefore $\tdelta'$ is not a symmetry of the system and the MSR description of the problem is not supersymmetric.

\newpage

\providecommand{\href}[2]{#2}\begingroup\raggedright\endgroup

\end{document}